\documentclass{WileyMSP-template}
\usepackage{graphicx} 
\usepackage{float}
\usepackage{geometry}
\usepackage[utf8]{inputenc}
\usepackage[T1]{fontenc} 
\usepackage{siunitx}
\usepackage{amssymb}
\usepackage[normalem]{ulem}
\usepackage{multibib}
\usepackage{braket}

\usepackage{color}
\definecolor{red}{rgb}{1,0,0}

\usepackage{amsmath}
\usepackage{authblk}
\usepackage[version=4]{mhchem}
\usepackage[english]{babel}
\makeatletter

\usepackage{hyperref}
\hypersetup{
    colorlinks = true,
    citecolor = {blue},
    urlcolor = {blue}
}
\newcommand{\mathunit}[1]{~\mathrm{#1}}

\DeclareUnicodeCharacter{2212}{-}
\DeclareUnicodeCharacter{2009}{ }
\DeclareUnicodeCharacter{03B4}{$\delta$}

\title{
Quantum sensing of weak ferromagnetism in a phase-separated Mn-doped CaFe$_3$O$_5$
}

\author[1]{Izidor Benedičič*}
\affil[1]{Jo\v{z}ef Stefan Institute, Jamova c. 39, 1000 Ljubljana, Slovenia}
\author[2]{J. Paul Attfield}
\affil[2]{Centre for Science at Extreme Conditions and School of Chemistry, University of Edinburgh, Mayfield Road, Edinburgh EH9 3FD, United Kingdom}
\author[1,3]{Denis Ar\v{c}on}
\affil[3]{Fakulteta za matematiko in fiziko, Univerza v Ljubljani, Jadranska c. 19, 1000 Ljubljana, Slovenia}
\date{*Email: izidor.benedicic@ijs.si}

\begin{document}
\pagestyle{fancy}

\maketitle

\author{Izidor Benedičič}
\author{J. Paul Attfield}
\author{Denis Arčon}

\begin{affiliations}
Izidor Benedičič\\
Address: Institut Jožef Stefan, Jamova c. 39, 1000 Ljubljana, Slovenia\\
Email Address: izidor.benedicic@ijs.si\\

J. Paul Attfield\\
Address: Centre for Science at Extreme Conditions and School of Chemistry, University of Edinburgh, Mayfield Road, Edinburgh EH9 3FD, United Kingdom\\

Denis Arčon\\
Address: Institut Jožef Stefan, Jamova c. 39, 1000 Ljubljana, Slovenia\\
Fakulteta za matematiko in fiziko, Univerza v Ljubljani, Jadranska c. 19, 1000 Ljubljana, Slovenia

\end{affiliations}

\begin{abstract}
\noindent
  Doping of transition metal oxides such as \ce{CaFe3O5} offers a controlled way to tune the interplay of charge, spin, and lattice degrees of freedom, yet local‑probe studies remain difficult because strong correlations and dynamic charge–spin fluctuations obscure fine spectroscopic features in powder samples.
  Here, we employ quantum magnetometry based on nitrogen-vacancy (NV) centers in nanodiamonds impressed into an Mn-doped \ce{CaFe3O5} powder pellet to probe static and dynamic magnetic fields across the weak ferromagnetic transition.
 The splitting and broadening of the optically detected magnetic resonance (ODMR) spectra exhibit an order-parameter-like increase by $\sim 15 \mathunit{MHz}$ upon cooling below the critical temperature, $T_{\rm c}$. 
 Concomitantly, the spin–lattice relaxation rate, $1/T_1$, exhibits a pronounced enhancement at $T_\mathrm{c}$, increasing approximately fivefold from its high-temperature value. Moreover, detailed lineshape fits of ODMR spectra together with the stretched-exponential NV magnetization recovery curves corroborate the proposed electronic phase segregation in charge-ordered and charge-averaged phases at the microscale. The presented study demonstrates the viability of using nanodiamonds as a platform for local magnetic probing of strongly correlated matter, including phenomena such as electronic phase separation. 
\end{abstract}


\twocolumn

\section{Introduction}

Materials with spin, charge, lattice, or orbital interactions all being simultaneously active often exhibit several competing electronic states. In such cases, the system may minimize energy by developing nanoscale inhomogeneities in electronic states, thereby breaking the symmetry of the governing Hamiltonian \cite{tokuraEmergent2017, morosanStrongly2012}. Prominent examples of such electronic complexity may be found across a broad range of transition metal oxides, including manganites \cite{moreoPhase1999,moritomoELectronic1999,miaoDirect2020} and cuprates \cite{bozinNeutron2000,langImaging2002,singer632002,campiFunctional2021}. In such electronic nanometer-scale structures, the emergent properties do not coincide with those associated with the parent electronically uniform states, as spectacularly demonstrated by colossal magneto-resistance in manganites, for example \cite{Dagotto2005,tokuraCritical2006}. \par
Recently, the competition between two electronic states has been discovered in lightly doped CaFe$_3$O$_5$ non-spinel ferrite with mixed Fe$^{2+}$/Fe$^{3+}$ charges \cite{hongLong2018, hongSubstitutional2021} (See Figure \ref{fig:basics}(a) for crystal structure and charge site assignment). The two competing phases are the charge-ordered (CO, with long-range \ce{Fe^{2+}}/\ce{Fe^{3+}} ordering) and charge-averaged (CA, where \ce{Fe} charge states remain mixed). The charge ordering transition temperature for CaFe$_{3-x}$$M_x$O$_5$ (here, $M=$ Co or Mn and $0.01\leq x \leq 0.1$) is at $320\mathunit{K}$, where the lattice strains drive part of the sample into CA and part into the CO phase. In the CO phase, Fe$^{2+}$/Fe$^{3+}$ charge ordering and Fe$^{2+}$ $t_{2g}$ orbital ordering arrangement establish the antiferromagnetic order of trimerons at $T_{\rm CO(m)} =  292$~K, with [1/2 0 0] propagation vector. The CA phase orders at $T_{\rm CA(m)} = 301$~K and adopts a different antiferromagnetic order with [0 0 0] propagation vector \cite{miltonProximate2024}. While both ordered phases are antiferromagnetic, magnetic susceptibility measurements show a small finite magnetization (approximately 0.05 $\mu_\mathrm{B}$ per unit cell) which was assigned to the CA phase \cite{cassidySingle2019}.
The coincidence of nearly degenerate energy scales for the two competing electronic states characterized by differing lattice strains and their long-range spin orders thus represents a novel route for materials with intrinsic electronic phase separation. \par
Spatially resolved techniques, such as scanning tunneling microscopy, magnetic force microscopy and transmission electron microscopy, have been indispensable for investigating electronic phase segregation at the nanoscale   \cite{langImaging2002,elbaggariNature2018,capuaDirect2006,duVisualization2015,zhouEvolution2015}. However, their application generally relies on the availability of high-quality single crystals or thin films, which are not accessible for the CaFe$_{3-x}M_x$O$_5$ non-spinel ferrites.
Nuclear magnetic resonance (NMR) is another established local-probe method to study electronic phase separation in bulk transition metal oxides \cite{singer632002,allodiElectronic1997,hunt631999,klaussAntiferromagnetic2000}. However, NMR studies of local magnetism in CaFe$_{3-x}M_x$O$_5$ appear extremely challenging: low natural abundances of available NMR-active nuclei and their extremely short relaxation times driven by strong spin and charge fluctuations substantially degrade the signal, highlighting the need for complementary local probes.\par
In the past decade, nitrogen-vacancy (NV) centers in diamonds have become established as a versatile method for nanoscale sensing of static and dynamic magnetic fields \cite{phamMagnetic2011, liCritical2025}, electrical conductivity \cite{ariyaratneNanoscale2018,kolkowitzProbing2015}, electric currents \cite{changNanoscale2017}, and temperature \cite{kucskoNanometrescale2013,tetienneScanning2016,petriniNanodiamond2022}. 
However, its use in strongly inhomogeneous systems has been limited so far, partly due to difficulties in ensuring good contact between the sample and the diamond sensor. \par
Here, we report on quantum sensing with nitrogen-vacancy centers in nanodiamonds to observe the emergence of weak ferromagnetism and electronic phase separation in Mn-doped \ce{CaFe3O5} (Mn-\ce{CaFe3O5} for short) by measuring the local stray magnetic field.
We use a novel, inexpensive method for sample preparation, utilizing commercially available nanodiamonds, which are pressed into a pelletized sample. 
This method facilitates good contact between the NV sensor and the sample while at the same time avoiding the problem of nanodiamond aggregation. Hence, it provides access to information about electronic inhomogeneities within the Mn-\ce{CaFe3O5} sample. Both optically detected magnetic resonance (ODMR) and the spin-lattice relaxation rates of NV centers are consistent with phase segregation into CA and CO phases. Our study provides unique local probe insight into the electronic phase segregation phenomena and highlights the potential of NV magnetometry in such studies.

\section{Results and discussion}

\begin{figure*}[htb]
	\centering
	\includegraphics[width=\textwidth]{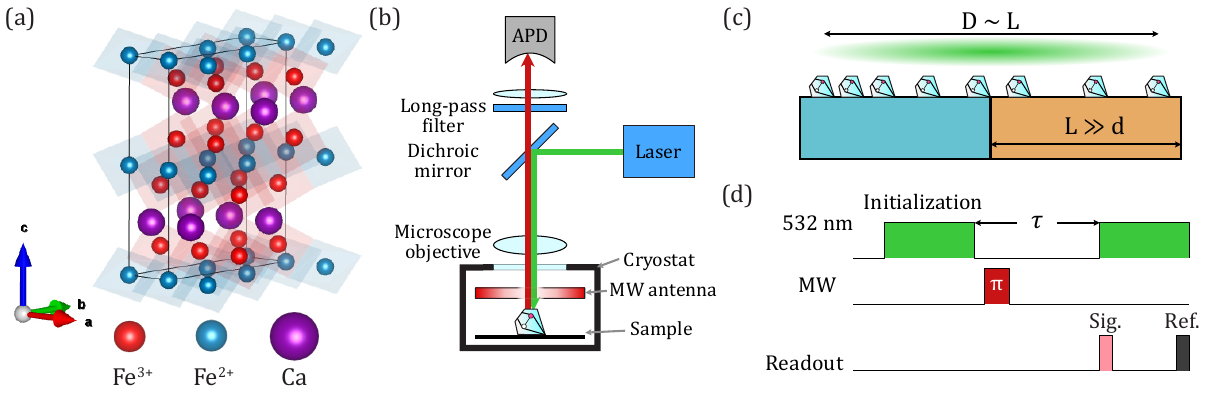}  
	\caption{Scheme of NV magnetometry experiments.
    (a) A three-dimensional model of \ce{CaFe3O5} crystal structure. In the charge-ordered phase, the structure comprises two inequivalent iron sites: \ce{Fe^{3+}} (red spheres) and \ce{Fe^{2+}} (blue spheres). In the charge-averaged phase, both sites have equal charge, \ce{Fe^{2.67+}}. Oxygen atoms are omitted for clarity. The structure was visualized using VESTA\cite{mommaVESTA2011}.
    (b) A schematic of the experimental setup for magnetometry and relaxometry using NV centers in nanodiamonds.
    (c) A cartoon depiction of nanodiamonds covering the sample. The size of magnetic domains $L$ is much larger than the average nanodiamond diameter $d$; the diameter of the laser beam $D$ is comparable with the domain size.
    (d) Pulse sequences for ODMR and $1/T_1$ measurements. The first laser pulse polarizes the population of NV centers in the $m_s = 0$ spin state. A resonant microwave pulse drives the transitions between the $m_s = 0$ and $m_s = \pm 1$ spin states, modifying their population. The second laser pulse reads out the resulting spin-state population through spin-dependent fluorescence. For the ODMR measurement, we sweep the frequency at a fixed delay pulse length $\tau$; for the $1/T_1$ measurement, we sweep $\tau$ at a fixed resonant microwave frequency.
    }
	\label{fig:basics}
\end{figure*}

In this study, we probe the magnetic transition in a powder sample of Mn-\ce{CaFe3O5} using a confocal microscopy setup, illustrated in Figure \ref{fig:basics}(b) and described in more detail in Ref. \cite{benedicicSpinlattice2025}. We prepared the sample by first compressing the Mn-\ce{CaFe3O5} powder in a pellet using a hydraulic press and then drop-casting the dispersion of nanodiamonds in deionized water. Before measurements, the pellets with deposited nanodiamonds were again compressed with the same conditions as before to ensure the nanodiamonds are in good contact with the pellet surface.
The nanodiamond used in this experiment had a nominal diameter of $d= 40 \mathunit{nm}$, much smaller than a typical magnetic domain, which were in previous works estimated to be at least $200 \mathunit{nm}$ \cite{hongLong2018}. The NV centers within each nanodiamond are thus primarily sensitive to local fields within a single magnetic domain, as illustrated in Figure \ref{fig:basics}(c).
\par 
The experimental pulse protocol depicted in Figure \ref{fig:basics} (d) consists of two $532 \mathunit{nm}$ laser pulses of $20 \mathunit{\mu s}$ duration for optical spin initialization and readout, separated by a delay $\tau$. Immediately after the first pulse, a microwave (MW) $\pi$ pulse is applied to drive the transition between the NV $m_s = 0 \leftrightarrow m_s = \pm 1$ states. For ODMR measurement, we sweep the microwave frequency while keeping $\tau$ constant. For spin-lattice relaxometry measurement, we sweep the delay $\tau$ at a fixed microwave frequency, and include a subsequent measurement without the MW pulse for normalization and noise cancellation \cite{bucherQuantum2019,mrozekLongitudinal2015}.\par

\begin{figure*}[ht]
	\centering
	\includegraphics[width=\textwidth]{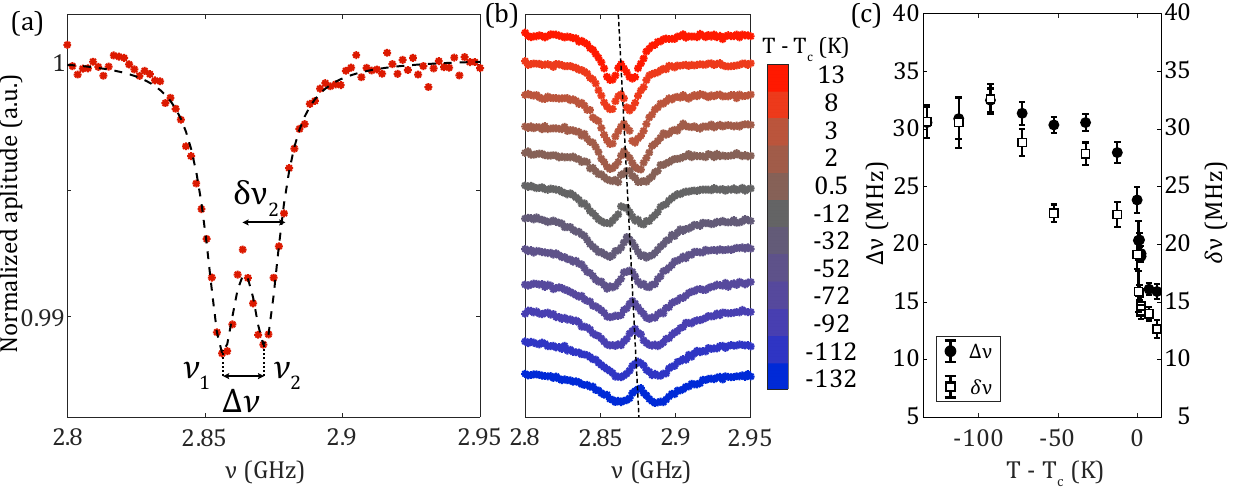}  
    	\caption{Temperature dependence of NV ODMR spectra.
        (a) Representative high-temperature ODMR spectrum of nanodiamonds implanted into a Mn-\ce{CaFe3O5} pellet, revealing the characteristic two-dip structure. The data are fitted with a sum of two Lorentzian functions (black dashed curve), yielding the resonance frequencies  $\nu_{1,2}$, the splitting $\Delta\nu$, and the spectral width $\delta\nu_{1,2}$, defined as the full width at half maximum (FWHM) of the individual Lorentzians. When $B=0$, the splitting reduces to the zero-field splitting parameter $E$.
        (b) A plot of ODMR spectra from the highest (red) to the lowest (blue) temperature. Below $T_\mathrm{c}$, increased splitting and broadening can be observed. The monotonic shift of the average spectral frequency (black dashed line) is a well-known intrinsic effect; we find $dD/dT = -78  \mathunit{kHz/K}$, consistent with values previously reported in the literature \cite{acostaTemperature2010}.
        (c) Temperature dependence of splitting $\Delta\nu$ (black circles) and spectral width $\delta \nu$ (white squares) as a function of temperature. The transition temperature $T_\mathrm{c}$ was determined from fitting an arctangent function to $\Delta\nu(T)$ -- see Supporting Information for more details. The error bars reflect the fitting uncertainty.
        }
	\label{fig:spectra}
\end{figure*}

The temperature dependencies of ODMR spectra and spin-lattice relaxation times of  Mn-\ce{CaFe3O5} were studied for several locations on the pelletized sample.
A typical NV-ODMR spectrum measured above the weak ferromagnetic transition temperature, shown in Figure \ref{fig:spectra} (a), comprises of two relatively broad absorption peaks. This is characteristic of nanodiamond ensembles and reflects the splitting of the $m_s = \pm1$ states in zero applied magnetic field conditions \cite{benedicicSpinlattice2025}. The ODMR spectrum can be fitted by a sum of two Lorentzian functions of approximately equal intensity, peaked at the frequencies $\nu_1$ and $\nu_2$, with respective full width at half maximum (FWHM) $\delta\nu_{1,2}$. High-temperature ODMR spectra already exhibit significant splitting $\Delta\nu=  \nu_2 - \nu_1 = 14 \pm 2 ~\mathrm{MHz} $ and a Lorentzian lineshape broadening of $\delta\nu_{1,2} = 15 \pm 3 ~\mathrm{MHz}$, consistent with the values previously reported for nanodiamonds deposited on glass \cite{benedicicSpinlattice2025}.\par
In the high-temperature phase of Mn-\ce{CaFe3O5}, the ODMR spectra [bright red colors in Figure \ref{fig:spectra}(b)] gradually shift towards higher frequencies with decreasing temperatures, which is a known phenomenon attributed to the effect of electron-phonon coupling on the zero-field splitting parameter at high temperatures \cite{acostaTemperature2010,chenTemperature2011,dohertyTemperature2014}. Both splitting $\Delta\nu$ as well as spectral width $\delta\nu_{1,2}$ are temperature-independent for $T>T_{\rm c}$.
However, as the sample temperature decreases below the critical temperature $T_{\rm c}$, the ODMR spectra [dark blue colors in Figure \ref{fig:spectra} (b)] change significantly.
Both the splitting and the linewidth of ODMR spectra dramatically increase upon cooling through the critical temperature.

The splitting extracted from the ODMR fits nearly doubles to $\Delta\nu= 30.6 \mathunit{MHz}$ at the lowest temperature. Concomitantly, the linewidth of an individual Lorentz peak increases to $\delta \nu= 31.3 \mathunit{MHz}$. The temperature dependencies of $\Delta\nu(T)$ and $\delta \nu(T)$, plotted in Figs. \ref{fig:spectra} (c) show an order-parameter-like dependence, which demonstrates the sensitivity of NV centers in nanodiamonds to magnetic fields from Mn-\ce{CaFe3O5}.
We observe such a transition at every measured location, though with some variation in the observed transition temperatures. These variations are discussed in detail in the Supporting Information.
We thus conclude that the observed changes to the ODMR spectra are due to additional static magnetic fields emerging at the weak ferromagnetic phase transition of Mn-\ce{CaFe3O5}, previously reported at $T_{\rm CA(m)} = 301$~K \cite{miltonProximate2024}.\par

By measuring the spin-lattice relaxation rate, we can also probe dynamic magnetic fields, obtaining information about magnetic spectral density across the transition.
In general, the relaxation rate of an NV center due to the fluctuations of magnetic fields can be written as
\begin{equation}\label{eq:relaxRateGeneral}
     \frac{1}{T_1} = \int \gamma^2 \langle B^2 \rangle S (\omega, T) F(\omega ) d\omega,
\end{equation}
where $\gamma$ is the gyromagnetic ratio of the NV center, $\sqrt{\langle B^2 \rangle }$ the effective magnetic field density at the position of the NV center, $S(\omega)$ is the spectral density of magnetic noise, and $F(\omega)$ is the filter function, which depends on the utilized pulse sequence. For an inversion-recovery $1/T_1$ measurement, the filter function is a narrow window centered at the ODMR frequency $\omega \approx 2\pi \cdot 2.87 \mathunit{GHz}$ at room temperature \cite{schafer-nolteTracking2014}.

\begin{figure}[ht]
	\centering
	\includegraphics[width=0.5\textwidth]{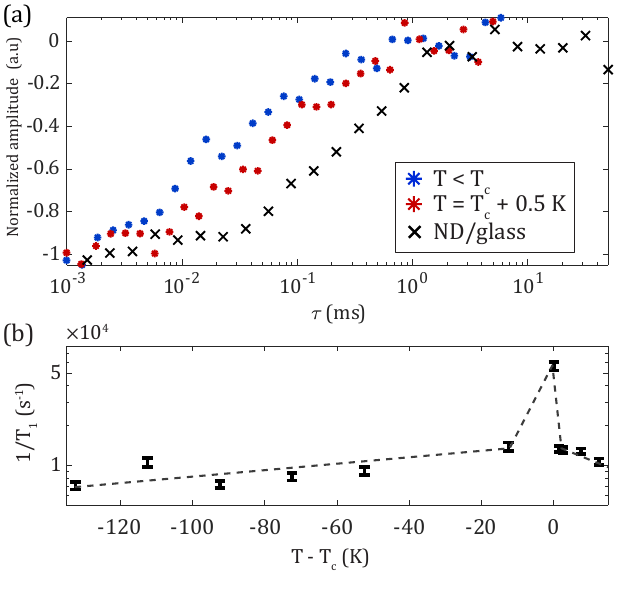}  
	\caption{NV relaxometry. (a) Normalized NV magnetization relaxation curves well below $T_\mathrm{c}$ ($T-T_\mathrm{c} = -132 \mathunit{K}$, blue symbols) and close to the transition temperature ($T-T_\mathrm{c} = 0.5 \mathunit{K}$, red symbols). NV spin-lattice relaxation rate, $1/T_1$, enhancement at $T_\mathrm{c}$ can be seen as the translation of the relaxation curve to the shorter times.
    Comparison with a magnetization relaxation curve for nanodiamonds deposited on glass (black crosses), taken from Ref. \cite{benedicicSpinlattice2025}, illustrates a significant enhancement in the relaxation rate when nanodiamonds are in contact with magnetic Mn-\ce{CaFe3O5}.
    (b) Temperature dependence of the NV spin-lattice relaxation rate measured on a Mn-\ce{CaFe3O5} pellet. The relaxation rate is strongly enhanced in the vicinity of $T_\mathrm{c}$. The gray dashed line is a guide to the eye.
    }
	\label{fig:t1}
\end{figure}

We measured the NV spin-lattice relaxation rate, $1/T_1$, at different locations on the sample and, in all cases, observed qualitatively the same behavior. Two NV magnetization relaxation curves taken on the same spot at $T<T_{\rm c}$ and at $T\approx T_{\rm c}$ are compared in Figure \ref{fig:t1}(a) together with a comparable NV magnetization relaxation curve of nanodiamonds deposited on a nonmagnetic glass substrate. The presence of the Mn-\ce{CaFe3O5} sample dramatically decreases the relaxation time.
A single-exponential model does not satisfactorily fit the measured relaxation curves (see Supporting information for details), which is reminiscent of our previous study of nanodiamonds deposited on a glass substrate \cite{benedicicSpinlattice2025}. We thus fit the NV magnetization recoveries using a stretched-exponential function,
\begin{equation}\label{eq:stretchedExp}
    A = A_0 e^{ - \left( \tau/T_1 \right)^\alpha },
\end{equation}
where $\alpha < 1$ is  the stretching exponent. The stretched-exponential relaxation reflects the fact that the measured signal originates from many NV centers within the ensemble of indented nanodiamonds, yielding a broad distribution of local relaxation times \cite{johnstonStretched2006}. 
An alternative explanation for such non-exponential recovery has been recently put forward in Ref. \cite{walsh_method_2026}. When polarization inefficiency within an ensemble is important, a two-state model can be better suited for fitting the relaxation curves than a stretched exponential model. We investigate the applicability of this method in the Supporting Information. Here, the choice to use a stretched exponential function still seems better justified: the signal originates from many nanodiamonds with different intrinsic relaxation rates, experiencing a broad distribution of magnetic fields from the sample.
\par
The temperature dependence of  $1/T_1$ is plotted in Figure \ref{fig:t1}(b). Above $T_\mathrm{c}$, the relaxation rate is on average $1/T_1 = 12000 \mathunit{s^{-1}} \pm 1400 \mathunit{s^{-1}}$. This is almost twice the intrinsic relaxation rate of the same nanodiamonds deposited on a glass substrate, $(1/T_1)_\mathrm{int} = 6250 \mathunit{s^{-1}} \pm  1000 \mathunit{s^{-1}}$, signifying that the NV centers experience substantial magnetic fluctuations from the sample.
Upon approaching $T_\mathrm{c}$ from above, the relaxation rate is pronouncedly enhanced by approximately a factor of 5 and reaches a maximum of $1/T_1 = 56 000\mathunit{s^{-1}} \pm 4000 \mathunit{s^{-1}}$ at $T=T_{\rm c}$. Such enhancement in $1/T_1$ points to an increase in the magnetic fluctuation spectrum at $\nu \approx 2.87 \mathunit{GHz}$ and is consistent with the onset of a magnetic transition.
Below $T_\mathrm{c}$, the relaxation rate is again suppressed to $1/T_1 = 8300 \mathunit{s^{-1}} \pm 500 \mathunit{s^{-1}}$, slightly lower than in the high-temperature regime. The observed dependence of $1/T_1$ reflects changes in the magnetic noise spectral density at $\nu \approx 2.87 \mathunit{GHz}$ and is fully consistent with the magnetic ordering in a powder sample, where the direction of local magnetic fields is random with respect to the NV axis.
The stretching exponent away from the transition is found to be $\alpha = 0.62 \pm 0.06$, consistent with the previous study of nanodiamond ensembles \cite{benedicicSpinlattice2025}. Close to $T_\mathrm{c}$, $\alpha$ decreases even further to $\alpha \approx 0.4$, consistent with broadening of the distribution of local relaxation rates for individual NV centers at the transition.

Having established that we successfully detect a weak ferromagnetic transition using NV centers in nanodiamonds, we further analyze the measured ODMR spectra to obtain quantitative information about the investigated magnetic phase.
A common approach in the weak magnetic field regime is to assume the spectrum can be described by a pair of Lorentzian curves, with their centres functions of temperature, magnetic field, electric field and strain:
\begin{equation}
    \nu_{1,2} = \nu_{1,2} ( T, \vec{B}, \vec{E}, \epsilon).
\end{equation}
However, constructing an effective model for an ensemble of NV centers in nanodiamonds is challenging, because each nanodiamond is subject to different mechanical strain \cite{awadallahSpinstrain2023} and local electric field. Both can significantly affect the overall shape of the ODMR spectrum \cite{mittigaImaging2018}. In addition, the crystal lattice axes of individual nanodiamonds in an ensemble are not aligned, resulting in a distribution of spectral line shapes even in a homogeneous magnetic field. Finally, the magnetic field of the powder sample cannot be assumed to be perfectly homogeneous within the experimentally illuminated spot ($d_\mathrm{laser} \approx 15 \mathunit{\mu m }$).\par
To obtain quantitative information from the measured spectra, we instead use a minimal assumptions approach, where we write down a microscopic Hamiltonian that includes Zeeman coupling, spin-strain coupling, and coupling with the electric field \cite{udvarhelyiSpinstrain2018, zhuSimulation2023}. We then calculate the NV ensemble ODMR spectrum by diagonalizing the Hamiltonian and summing the resulting spectra over a large (N = 2000) number of nanodiamonds. The details of the calculation are presented in the Supporting Information.\par

\begin{figure}[ht]
	\centering
	\includegraphics[width=0.5\textwidth]{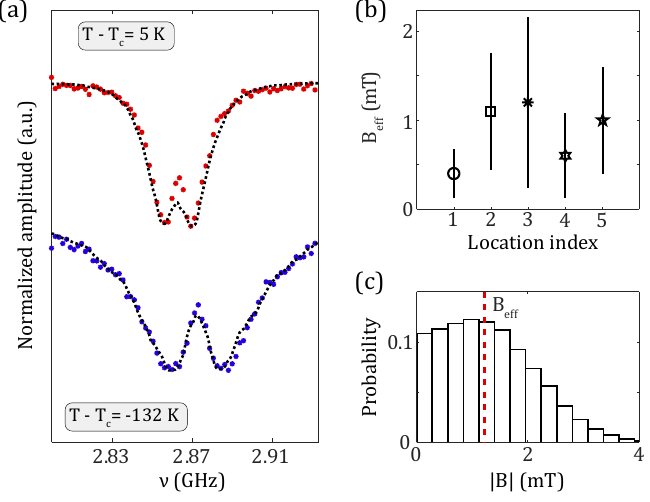}  
	\caption{Numerical simulations of the ODMR spectra. (a) Experimental ODMR spectra recorded above (red symbols) and below (blue symbols) the weak ferromagnetic transition of Mn-\ce{CaFe3O5}. The overlaid black dashed lines correspond to numerical simulations. For fitting the high-temperature spectrum, we only included the uncompensated Earth's magnetic field of $B_\mathrm{earth} = 45 \mathunit{\mu T}$. For fitting the low-temperature ODMR spectrum, we had to include an additional field on the order of $B \sim 1 \mathunit{mT}$ from the Mn-\ce{CaFe3O5} pellet.
    (b) Comparison of the fitted parameters obtained for several investigated locations on the Mn-\ce{CaFe3O5} pellet. The symbol position denotes the effective field $B_\mathrm{eff}$ at the NV site, and the vertical line spans the interval $B_\mathrm{eff} \pm \sigma_B$ providing a measure of the field-distribution width.
    (c) Distribution of magnetic field magnitudes for simulating the low-temperature spectrum in panel (a). The red dashed line indicates the effective magnetic field $B_\mathrm{eff}$.
    }
	\label{fig:fitting}
\end{figure}

We first sought to determine the effective electric field and strain in the high-temperature phase. Since the ensemble in question contains multiple nanodiamonds with random relative crystallographic orientations, the field is averaged over several independent spatial directions, and the information obtained from such fitting is reduced to a single scalar, the effective value of the electric field $E_\mathrm{eff}$. The best fit, shown in Figure \ref{fig:fitting} (a), was obtained with an effective electric field of $E_\mathrm{eff} = 50 \mathunit{MV/m}$. This value is unusually high; however, we note that within our approach, it is not possible to decouple the strain and electric-field contributions, since their Hamiltonians are symmetry equivalent \cite{awadallahSpinstrain2023}. This does not affect our main findings, since the aim of the work is not to determine internal strain and electric fields within nanodiamonds but to reconstruct the magnetic fields emerging at the weak ferromagnetic transition .\par
When simulating the spectra in the ferromagnetic phase, we assume both strain and electric field within the nanodiamonds remain unchanged, and the only difference is the additional magnetic field from the Mn-\ce{CaFe3O5} sample. 
The local magnetic field at the location of each nanodiamond is drawn from a normal distribution with standard deviation of $\sigma_B$ and mean at $B_\mathrm{eff}$. Such a choice is justified because the signal is averaged over a large ensemble of nanodiamonds covering different domains and domain walls, as well as grains with different crystallographic orientations.
In the low-temperature regime, the strong tails of the ODMR spectra are especially very pronounced. In fact, we obtain satisfactory fits only when the standard deviation of the field values becomes comparable to the effective mean magnitude $\sigma_B \approx B_\mathrm{eff}$. The best fit to the low-temperature spectra from Figure \ref{fig:spectra} is replotted in Figure \ref{fig:fitting}(a), with fit parameters determined to be $B_\mathrm{eff} = 1.2 \pm 0.1 \mathunit{mT}$ and $\sigma_B = 0.96 \mathunit{mT}$. The fit parameters collected for several different measured areas are summarized in Figure \ref{fig:fitting}(b). The magnitude of $B_\mathrm{eff}$ is consistent with the values expected for NV centers in proximity to the surface of a ferromagnet \cite{thielProbing2019, fincoSingle2023, schmid-lorchRelaxometry2015}.\par
The fit of ODMR spectrum yields a distribution of magnetic fields plotted in Fig. \ref{fig:fitting} (c) and shows that a significant fraction of NV centers experiences a very low magnetic field, while the high-field region follows the expected exponential drop-off. Additionally, the fitting seems to demand a notable fraction of nanodiamonds in a relatively large magnetic field, $|B|>2 \mathunit{mT}$. Such a distribution seems implausible, given that the material in question is a weak ferromagnet.

Combined with the prior knowledge of the phase separation in Mn-\ce{CaFe3O5}, a substantial weight at $B=0$ and the broad lineshapes stimulated us to approach the NV ODMR fitting with an alternative model, based on the electronic phase segregation into a weakly ferromagnetic CA and antiferromagnetic CO phases, observed in powder diffraction studies \cite{miltonProximate2024}. To justify this alternative fitting strategy, we first note that the laser spot size, $d_\mathrm{laser} \approx 15 \mathunit{\mu m}$, is comparable to the typical size of a magnetic domain, $d_\mathrm{domain} \approx 10 \mathunit{\mu m}$. 
The electron microscopy images of the pelletized sample (Figure S1 in Supporting information) confirm that the pellet is generally uniformly covered with nanodiamonds over a scale of micrometers. Therefore, multiple-component ODMR contributions cannot solely be attributed to the agglomeration of nanodiamonds.
We can thus postulate that the observed low-temperature spectra are composed of two distinct Mn-\ce{CaFe3O5} magnetic contributions. The first originates from NV centers in nanodiamonds deposited on a weakly ferromagnetic domain, exhibiting strong splitting and broadening, while the second comes from NV centers coupled to an antiferromagnetic part of the sample. Because stray fields within an antiferromagnetic domain are very small ($|B|<100\mathunit{\mu T}$), the NV centers within these domains remain virtually unaffected by the onset of the magnetic order. The total measured spectrum is then a linear combination of these two contributions:

\begin{equation}
    L(\nu) = A_1L_\mathrm{AFM}(\nu) + A_2L_\mathrm{WFM}(\nu),
\end{equation}
where $L_\mathrm{AFM/WFH}$ denotes the absorption spectrum originating from the NV centers deposited on the antiferromagnetic and the weakly ferromagnetic phases, respectively. The shape of $L_\mathrm{AFM/WFM}$ was calculated by setting $B=0$ and $|B| = 1.2 \mathunit{mT}, \sigma_B = 0.27 \mathunit{mT}$. The choice of magnetic field magnitude is based on the maximum $B_\mathrm{eff}$ in Fig. 4. (b). Within this model, such a choice is natural - the mean of the field distribution should be maximal when ODMR is measured within a single weakly ferromagnetic domain. The relative spectral fraction $\eta_{1,2} = 100\cdot A_{1,2}/(A_1+A_2)$ then reflects the relative abundance of each type of spectrum within the illuminated spot.\par
\begin{figure*}[h]
	\centering
	\includegraphics[width=\textwidth]{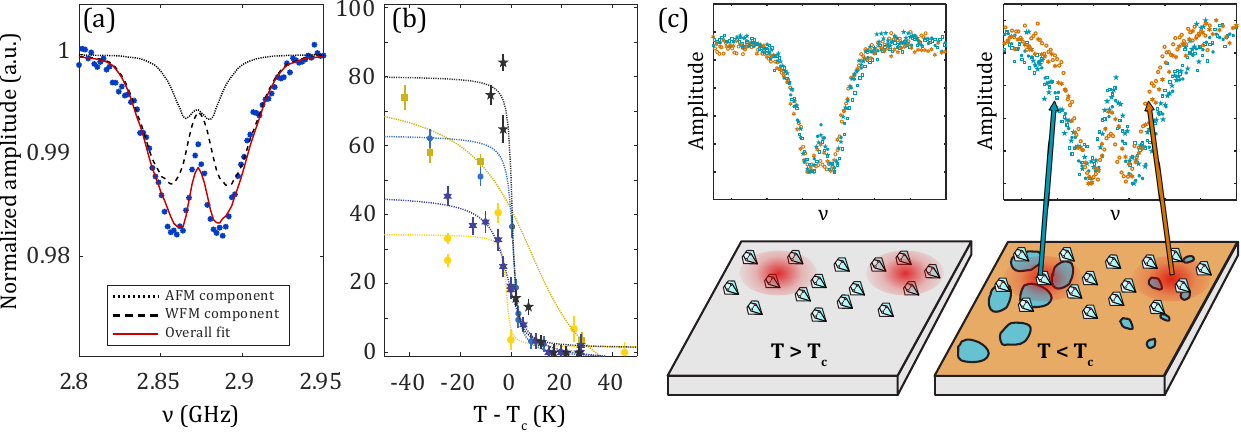}  
    \caption{Reconstruction of electronic phase fractions from numerical fitting of spectra.
    (a) Two-component fit of a low-temperature ($T-T_\mathrm{c}=-132\mathunit{K}$) ODMR spectrum of nanodiamonds deposited on a Mn-\ce{CaFe3O5} pellet. The dashed and the dotted black curves denote the two individual components contributing to the overall fit (solid red line).
    (b) Temperature dependence of the weak ferromagnetic spectral fraction $\eta_2$ for measurements at 5 locations on the sample. The dotted lines are guides to the eye.
    (c) Schematic illustration of the connection between ODMR spectral changes and electronic phase segregation. The spectra from locations with high remaining AFM fraction $\eta_1$ (1 and 4) are plotted in orange, the rest in cyan color. At high temperatures (left), all nanodiamonds probe an equivalent local environment, and intrinsic nanodiamond-to-nanodiamond differences dominate the observed small spectral variations. Below the magnetic transition, the sample electronically phase-segregates into weakly ferromagnetic (cyan) and antiferromagnetic (orange) domains. 
    Since the typical domain size is comparable to the laser beam size (shown as a red spot), the spectra can contain the signal from nanodiamonds covering both phases. The ODMR spectra develop additional broadening and splitting from the weakly ferromagnetic phase. The degree of additional broadening and splitting reflects the local fraction of the weakly ferromagnetic phase.
    }
	\label{fig:simulations}
\end{figure*}

Such a two-component fit, shown in Figure \ref{fig:simulations}(a), can correctly describe both the strong tails and the central maximum of the low-temperature ODMR spectra, and is statistically indistinguishable from the fit with a broad distribution (see Supporting Information for details). We further attempt to fit the two-component model to measured spectra in a wide range of temperatures. Plotting the relative fraction of the WFM spectral contribution $\eta_{2}$ for several measurement locations [Figure \ref{fig:simulations} (b)], we find a distribution of saturation values of $\eta_2$, ranging from $34 \% - 80\%$. Because of the uncertainty in the choice of $|B|$ for the WFM phase, $\eta_2$ should be regarded as an upper bound rather than an exact calculation of the fraction of nanodiamonds within the WFM phase.
The proposed origin of such observed spread is illustrated in Figure \ref{fig:simulations}(c), where we plot the ODMR spectra from all the measured locations: data from locations 1 and 4 are plotted in orange while the data from locations 2,3, and 5 are plotted in cyan. At $T>T_c$, all spectra are very similar. In the magnetically ordered phase, $T<T_\mathrm{c}$, the degree of spectral broadening at a given measurement location reflects the distribution of local magnetic fields and, by extension, the distribution of local electronic phases.\par

Previous synchrotron X-ray and neutron diffraction measurements on the samples from the same batch revealed that both the CO and CA phases coexist down to the lowest measured temperatures. The phase fraction of the CA phase, determined by XRD and neutron diffraction measurements, varies between $\approx$ 25 \% - 50 \%, depending on the thermal history of the sample \cite{hongLong2018,miltonProximate2024}. 
Our measurements are thus consistent with the picture where local electronic phase inhomogeneity gives rise to varying local magnetic fields, which is in turn reflected in variations in magnetic broadening and splitting of the ensemble ODMR spectra. We note, however, that the extracted spectral fraction cannot be simply identified with the phase fraction, and measurements within a single domain or wide-field imaging across multiple domains would be required to definitely establish the link between the two quantities.

\section{Conclusions}

We used NV centers in nanodiamonds as quantum sensors for the detection of weak magnetic fields produced by a powder sample of Mn-doped \ce{CaFe3O5}. We successfully observed the emergence of weak ferromagnetism by tracking the temperature dependence of ODMR spectra, which show additional broadening and splitting at the transition temperature. The spin-lattice relaxation rate $1/T_1$ increases significantly at the transition temperature, confirming that the spectral changes originate from the magnetic transition. Modeling the absorption spectra of NV center ensembles in nanodiamonds is consistent with a substantial portion of the NV centers experiencing very low magnetic fields. The ODMR spectrum of the NV ensemble can be described by a combination of magnetically broadened and zero-field spectra, with their ratios close to the phase fractions of the electronic phases found in Mn-doped \ce{CaFe3O5}, pointing to the scenario where the two types of spectra correspond to different magnetic orders of the two phases.\par
This study demonstrates the potential of NV magnetometry to study electronic phase separation phenomena in correlated electron materials by detecting spatial variations of magnetic properties. As such, it represents an alternative, exceptionally sensitive method in cases where other local probes, such as NMR or NQR, are impractical, when access to quality single crystals is limited, or when the differences in magnetization of individual phases are too small to be reliably observed by different methods.

\section*{Experimental section}

\textbf{Sample synthesis.} The polycrystalline sample of \ce{CaFe_{2.99}Mn_{0.01}O5} was synthesized as part of the work reported in Ref. \cite{hongSubstitutional2021}. 

\textbf{Nanodiamond deposition.} As magnetic field sensors, we used fluorescent nanodiamonds with nominal diameter $d=40\mathunit{nm}$ purchased from Ad\'amas Nanotechnologies. The diamond particles were produced by milling synthetic diamond manufactured using high-pressure high-temperature (HPHT) synthesis. They contain approximately 100~ppm of nitrogen substitutions and were irradiated with $2-3 \mathunit{MeV}$ electrons with a fluence of $1\times10^{19} \mathunit{e/cm^2}$. The obtained samples were annealed at $850^{\circ} \mathrm{C}$ for 2 hours under vacuum \cite{shenderovaCommercial2017,nunnBrilliant2019}. The final concentration of NV centers was 1.5~ppm, corresponding to an average of 14 NV centers per nanodiamond. \par
The samples were prepared for measurements by first compressing the \ce{CaFe3O5} powder into a pellet using a hydraulic press ($F = 50 \mathunit{kN}, p \approx 1 \mathunit{GPa}$) for 2 minutes. The dimensions of the pellet were $d=8\mathunit{mm}$ in diameter and $h=1.6\mathunit{mm}$ in thickness. The nanodiamonds were dispersed in deionized water before depositing two $V=5\mathunit{\mu L}$ droplets on a \ce{CaFe3O5} pellet. The composite samples were left for an additional 2 hours until the water evaporated, and then compressed again, using the same pressure and time, to ensure good contact between the nanodiamonds and the sample. To assess the nanodiamond coverage, we imaged the sample surface with a scanning electron microscope (SEM). The representative images are shown in the Supporting information.

\textbf{Experimental setup.} The NV magnetometry and relaxometry measurements were performed with a home-built confocal microscopy setup described in Benedičič \textit{et al.} \cite{benedicicSpinlattice2025}. All measurements were performed in a Lake Shore ST-500 continuous-flow cryostat cooled with liquid helium with a temperature stability better than $\Delta T =\pm 0.1\mathunit{K}$. The samples were mounted on a copper cold finger with a resistive heater, enabling measurements between room temperature and 5 K. A green laser with a 532 nm wavelength (Coherent Verdi G5) was attenuated to $ P = 3.2 \mathunit{mW}$ and focused using a microscope objective (Nikon CFI S Plan Fluor ELWD 20XC). We estimate the beam waist diameter at the sample to be $d_\mathrm{laser} = 15 \mathunit{\mu m}$, measured by making an image of the beam on a reference sample with distance markers.
The fluorescent light was collected with the same objective lens, passed through a dichroic mirror (Thorlabs DMLP605) and a laser-line filter (Thorlabs NF533-17) to remove the residual green light, and detected with a variable-gain avalanche photodiode (Thorlabs APD410A). The electrical signal was collected with an oscilloscope (Keysight DSOX1204G).
For microwave (MW) pulse delivery, we used a home-built omega-shaped copper stripline resonator on a printed circuit board (PCB), with the diameter of $d \sim 3 ~\mathrm{mm}$.
The pulses were generated with a radiofrequency signal generator (Stanford Research Systems SG384). The laser and microwave pulses were programmed and synchronized using a SpinCore PulseBlasterESR-PRO. All measurements reported here were performed in a zero external magnetic field.\par

\textbf{NV magnetometry and relaxometry.}
We measured the optically detected magnetic resonance signal using a standard pulsed sequence \cite{benedicicSpinlattice2025}, schematically illustrated in Figure \ref{fig:basics} (d). The sequence begins with a $20 \mathunit{\mu s}$ long laser pulse, polarizing the NV centers to an $m_s = 0$ state. We then apply a microwave $\pi$ pulse for the duration of $\tau_{\pi} = 130 \mathunit{ns}$. The $\pi$ pulse only excites the NV centers within the $1/2\tau_{\pi} = 4 \mathunit{MHz}$ window around the set MW frequency, much narrower than the full width of the ODMR spectrum, therefore effectively selecting a sub-ensemble of NV centers. After $t_\mathrm{dark} = 2 \mathunit{\mu s}$, another laser pulse is applied to read the NV fluorescence and determine the spin polarization. The initial signal of the photodiode was integrated within the integration window of $t_\mathrm{int} = 3.1 ~\mathunit{\mu s}$ and normalized to the signal value at the end of the readout pulse. We repeated this measurement $N_\mathrm{rep} = 12000$ times for each microwave frequency in the interval between $\nu = 2.80-2.95 \mathunit{GHz}$.

The spin-lattice relaxation rate, $1/T_1$, was measured at a frequency of the lower-frequency peak, fitted to the ODMR spectrum. The pulse sequence for the $1/T_1$ measurements begins with a $20 \mathunit{\mu s}$ long laser pulse, followed by a microwave $\pi$ pulse at the frequency $\nu_1$.
The system is then left to evolve for a variable time $\tau$, after which another laser pulse is applied to read the NV fluorescence and determine the residual spin polarization. The photodiode signal was integrated within a time window of $100\mathunit{ns}$ to obtain the initial signal (S)  and normalized to the reference signal value after a long time (R). The normalized NV polarization signal is then calculated as $A_\mathrm{MW} = (R-S)/R$.
This sequence is immediately followed by a control sequence with the same parameters, except without the MW pulse, yielding $A_\mathrm{0}$. In the final stage, the two results are subtracted to obtain a signal $A_\mathrm{CMR} = A_\mathrm{MW} - A_\mathrm{0}$, which is proportional to the polarization of only those NV centers that were excited by the MW pulse \cite{jarmolaLongitudinal2015,mrozekLongitudinal2015}.

\textbf{Numerical simulations}
We simulated the ODMR spectra of NV centers by direct diagonalization. The Hamiltonian describing the individual NV center can be written as
\begin{equation}
    H = H_\mathrm{zfs} + H_B + H_E + H_\epsilon + H_\mathrm{nuc},
\end{equation}
where $H_\mathrm{zfs}$ describes zero-field splitting, $H_E$ captures the interaction of electron spin with electric field, $H_B$ captures the interaction of electron spin with magnetic field, $H_\epsilon$ the interaction with lattice strain and $H_\mathrm{nuc}$ captures the nuclear effects. We included all the mentioned terms, diagonalized the Hamiltonian, and calculated the eigenfunctions. The MW absorption spectrum was then calculated by calculating the transition probability between initial and final states using an interaction Hamiltonian  Lorentzian broadening of each allowed transition by a constant factor $\Gamma = 1 \mathunit{MHz}$. The parameters of the Hamiltonian are listed in the Supporting information.\par
To describe an ensemble spectrum, we assigned a value of lattice strain, electric field and magnetic field to each nanodiamond, according to a normal distribution:
\begin{equation}
     P_A = \frac{1}{\sigma_A \sqrt{2\pi}} \exp \left( -\frac{(A-\mu_{A})^2}{2\sigma_A^2} \right),
\end{equation}
where $P_A$ is the probability density, $\mu_A$ is the mean value and $\sigma_A$ is the standard deviation of the distribution of quantity $A$. We then performed the calculation described above for $N=2000$ nanodiamonds and summed the individual absorption spectra. We also took into account that each nanodiamond has on average 14 NV centers, randomly oriented along one of the four bond directions. The absorption strength was rescaled to match the experimental contrast.

\section*{Acknowledgements.}
The authors gratefully acknowledge dr. Viliam Branislav Hakala and dr. Ka H. (Jacky) Hong for synthesis of the Mn-doped \ce{CaFe3O5} powder, dr. Gregor Kladnik for performing scanning electron microscopy imaging of the sample surface, and dr. Polona Umek for her help with the sample preparation.
The authors also thank dr. Mariusz Mrózek and prof. Adam Wojciechowski from Jagiellonian University for insightful discussions.
IB acknowledges the financial support from the Slovenian Research and Innovation Agency (ARIS) the P1-0125 research program.
DA acknowledges the financial support from the Slovenian Research and Innovation Agency (ARIS) through the J1-70021 research grant, P1-0125 research program and UL VIP project (KTTK21) under contract no. SN-ZRD/22-27/510.
JPA acknowledges EPSRC (UK) for support.

\onecolumn
\bibliographystyle{unsrt}
\bibliography{references.bib}

\begin{thebibliography}{10}

\bibitem{tokuraEmergent2017}
Yoshinori Tokura, Masashi Kawasaki, and Naoto Nagaosa.
\newblock Emergent functions of quantum materials.
\newblock {\em Nature Physics}, 13(11):1056--1068, November 2017.

\bibitem{morosanStrongly2012}
Emilia Morosan, Douglas Natelson, Andriy~H. Nevidomskyy, and Qimiao Si.
\newblock Strongly {Correlated} {Materials}.
\newblock {\em Advanced Materials}, 24(36):4896--4923, September 2012.

\bibitem{moreoPhase1999}
Adriana Moreo, Seiji Yunoki, and Elbio Dagotto.
\newblock Phase {Separation} {Scenario} for {Manganese} {Oxides} and {Related} {Materials}.
\newblock {\em Science}, 283(5410):2034--2040, March 1999.

\bibitem{moritomoELectronic1999}
Y.~Moritomo, A.~Machida, S.~Mori, N.~Yamamoto, and A.~Nakamura.
\newblock Electronic phase diagram and phase separation in {Cr}-doped manganites.
\newblock {\em Physical Review B}, 60(13):9220--9223, October 1999.

\bibitem{miaoDirect2020}
Tian Miao, Lina Deng, Wenting Yang, Jinyang Ni, Changlin Zheng, Joanne Etheridge, Shasha Wang, Hao Liu, Hanxuan Lin, Yang Yu, Qian Shi, Peng Cai, Yinyan Zhu, Tieying Yang, Xingmin Zhang, Xingyu Gao, Chuanying Xi, Mingliang Tian, Xiaoshan Wu, Hongjun Xiang, Elbio Dagotto, Lifeng Yin, and Jian Shen.
\newblock Direct experimental evidence of physical origin of electronic phase separation in manganites.
\newblock {\em Proceedings of the National Academy of Sciences}, 117(13):7090--7094, March 2020.

\bibitem{bozinNeutron2000}
E~S Božin, G~H Kwei, H~Takagi, and S~J~L Billinge.
\newblock Neutron {Diffraction} {Evidence} of {Microscopic} {Charge} {Inhomogeneities} in the {CuO}$_{\textrm{2}}$ {Plane} of {Superconducting} {La}$_{\textrm{2-x}}${Sr}$_{\textrm{x}}${CuO}$_{\textrm{4}}$ (0{\textless}x{\textless}0.30).
\newblock {\em Physical Review Letters}, 84(25), 2000.

\bibitem{langImaging2002}
K.~M. Lang, V.~Madhavan, J.~E. Hoffman, E.~W. Hudson, H.~Eisaki, S.~Uchida, and J.~C. Davis.
\newblock Imaging the granular structure of high-{Tc} superconductivity in underdoped {Bi2Sr2CaCu2O8}+δ.
\newblock {\em Nature}, 415(6870):412--416, January 2002.

\bibitem{singer632002}
P.~M. Singer, A.~W. Hunt, and T.~Imai.
\newblock Cu$^{\textrm{63}}$ {NQR} {Evidence} for {Spatial} {Variation} of {Hole} {Concentration} in {La}$_{\textrm{2-x}}${Sr}$_{\textrm{x}}${CuO}$_{\textrm{4}}$.
\newblock {\em Physical Review Letters}, 88(4):047602, January 2002.

\bibitem{campiFunctional2021}
Gaetano Campi and Antonio Bianconi.
\newblock Functional {Nanoscale} {Phase} {Separation} and {Intertwined} {Order} in {Quantum} {Complex} {Materials}.
\newblock {\em Condensed Matter}, 6(4):40, November 2021.

\bibitem{Dagotto2005}
Elbio Dagotto.
\newblock Complexity in strongly correlated electronic systems.
\newblock {\em Science}, 309(5732):257--262, July 2005.
\newblock arXiv: cond-mat/0509041.

\bibitem{tokuraCritical2006}
Y~Tokura.
\newblock Critical features of colossal magnetoresistive manganites.
\newblock {\em Reports on Progress in Physics}, 69(3):797--851, March 2006.

\bibitem{hongLong2018}
Ka.~H. Hong, Angel~M. Arevalo-Lopez, James Cumby, Clemens Ritter, and J.~Paul Attfield.
\newblock Long range electronic phase separation in {CaFe}$_{\textrm{3}}${O}$_{\textrm{5}}$.
\newblock {\em Nature Communications}, 9(1):2975, July 2018.

\bibitem{hongSubstitutional2021}
Ka~H. Hong, Elena Solana-Madruga, Branislav~Viliam Hakala, Midori~Amano Patino, Pascal Manuel, Yuichi Shimakawa, and J.~Paul Attfield.
\newblock Substitutional tuning of electronic phase separation in {CaFe}$_{\textrm{3}}${O}$_{\textrm{5}}$.
\newblock {\em Physical Review Materials}, 5(2):024406, February 2021.

\bibitem{miltonProximate2024}
Michael~J. Milton, Branislav~V. Hakala, Ka~H. Hong, Maxim Avdeev, Chris~D. Ling, Brendan~J. Kennedy, Pascal Manuel, and J.~Paul Attfield.
\newblock Proximate electronic and magnetic phase transitions in {CaFe}$_{\textrm{3}}${O}$_{\textrm{5}}$.
\newblock {\em Physical Review Materials}, 8(12):124408, December 2024.

\bibitem{cassidySingle2019}
Simon~J. Cassidy, Fabio Orlandi, Pascal Manuel, and Simon~J. Clarke.
\newblock Single phase charge ordered stoichiometric {CaFe}$_{\textrm{3}}${O}$_{\textrm{5}}$ with commensurate and incommensurate trimeron ordering.
\newblock {\em Nature Communications}, 10(1):5475, December 2019.

\bibitem{elbaggariNature2018}
Ismail El~Baggari, Benjamin~H. Savitzky, Alemayehu~S. Admasu, Jaewook Kim, Sang-Wook Cheong, Robert Hovden, and Lena~F. Kourkoutis.
\newblock Nature and evolution of incommensurate charge order in manganites visualized with cryogenic scanning transmission electron microscopy.
\newblock {\em Proceedings of the National Academy of Sciences}, 115(7):1445--1450, February 2018.

\bibitem{capuaDirect2006}
R~Di Capua, C~A Perroni, V~Cataudella, F~Miletto Granozio, P~Perna, M~Salluzzo, U~Scotti~Di Uccio, and R~Vaglio.
\newblock Direct observation of spectroscopic inhomogeneities on {La}$_{\textrm{0.7}}$ {Sr}$_{\textrm{0.3}}$ {MnO}$_{\textrm{3}}$ thin films by scanning tunnelling spectroscopy.
\newblock {\em Journal of Physics: Condensed Matter}, 18(35):8195--8204, September 2006.

\bibitem{duVisualization2015}
Kai Du, Kai Zhang, Shuai Dong, Wengang Wei, Jian Shao, Jiebin Niu, Jinjie Chen, Yinyan Zhu, Hanxuan Lin, Xiaolu Yin, Sy-Hwang Liou, Lifeng Yin, and Jian Shen.
\newblock Visualization of a ferromagnetic metallic edge state in manganite strips.
\newblock {\em Nature Communications}, 6(1):6179, February 2015.

\bibitem{zhouEvolution2015}
Haibiao Zhou, Lingfei Wang, Yubin Hou, Zhen Huang, Qingyou Lu, and Wenbin Wu.
\newblock Evolution and control of the phase competition morphology in a manganite film.
\newblock {\em Nature Communications}, 6(1):8980, November 2015.

\bibitem{allodiElectronic1997}
G.~Allodi, R.~De~Renzi, G.~Guidi, F.~Licci, and M.~W. Pieper.
\newblock Electronic phase separation in lanthanum manganites: {Evidence} from 55 {Mn} {NMR}.
\newblock {\em Physical Review B}, 56(10):6036--6046, September 1997.

\bibitem{hunt631999}
A.~W. Hunt, P.~M. Singer, K.~R. Thurber, and T.~Imai.
\newblock Cu$^{\textrm{63}}$ {NQR} {Measurement} of {Stripe} {Order} {Parameter} in {La}$_{\textrm{2-x}}${Sr}$_{\textrm{x}}${CuO}$_{\textrm{4}}$.
\newblock {\em Physical Review Letters}, 82(21):4300--4303, May 1999.

\bibitem{klaussAntiferromagnetic2000}
H.-H. Klauss, W.~Wagener, M.~Hillberg, W.~Kopmann, H.~Walf, F.~J. Litterst, M.~Hücker, and B.~Büchner.
\newblock From {Antiferromagnetic} {Order} to {Static} {Magnetic} {Stripes}: {The} {Phase} {Diagram} of ({La}, {Eu})$_{\textrm{2-x}}${Sr}$_{\textrm{x}}${CuO}$_{\textrm{4}}$.
\newblock {\em Physical Review Letters}, 85(21):4590--4593, November 2000.

\bibitem{phamMagnetic2011}
L~M Pham, D~Le~Sage, P~L Stanwix, T~K Yeung, D~Glenn, A~Trifonov, P~Cappellaro, P~R Hemmer, M~D Lukin, H~Park, A~Yacoby, and R~L Walsworth.
\newblock Magnetic field imaging with nitrogen-vacancy ensembles.
\newblock {\em New Journal of Physics}, 13(4):045021, April 2011.

\bibitem{liCritical2025}
Yuxin Li, Zhe Ding, Chen Wang, Haoyu Sun, Zhousheng Chen, Pengfei Wang, Ya~Wang, Ming Gong, Hualing Zeng, Fazhan Shi, and Jiangfeng Du.
\newblock Critical fluctuations and noise spectra in two-dimensional {Fe}$_{\textrm{3}}${GeTe}$_{\textrm{2}}$ magnets.
\newblock {\em Nature Communications}, 16(1):8585, September 2025.

\bibitem{ariyaratneNanoscale2018}
Amila Ariyaratne, Dolev Bluvstein, Bryan~A. Myers, and Ania C.~Bleszynski Jayich.
\newblock Nanoscale electrical conductivity imaging using a nitrogen-vacancy center in diamond.
\newblock {\em Nature Communications}, 9(1):2406, June 2018.

\bibitem{kolkowitzProbing2015}
S.~Kolkowitz, A.~Safira, A.~A. High, R.~C. Devlin, S.~Choi, Q.~P. Unterreithmeier, D.~Patterson, A.~S. Zibrov, V.~E. Manucharyan, H.~Park, and M.~D. Lukin.
\newblock Probing {Johnson} noise and ballistic transport in normal metals with a single-spin qubit.
\newblock {\em Science}, 347(6226):1129--1132, March 2015.

\bibitem{changNanoscale2017}
K.~Chang, A.~Eichler, J.~Rhensius, L.~Lorenzelli, and C.~L. Degen.
\newblock Nanoscale {Imaging} of {Current} {Density} with a {Single}-{Spin} {Magnetometer}.
\newblock {\em Nano Letters}, 17(4):2367--2373, April 2017.

\bibitem{kucskoNanometrescale2013}
G.~Kucsko, P.~C. Maurer, N.~Y. Yao, M.~Kubo, H.~J. Noh, P.~K. Lo, H.~Park, and M.~D. Lukin.
\newblock Nanometre-scale thermometry in a living cell.
\newblock {\em Nature}, 500(7460):54--58, August 2013.

\bibitem{tetienneScanning2016}
Jean-Philippe Tetienne, Alain Lombard, David~A. Simpson, Cameron Ritchie, Jianing Lu, Paul Mulvaney, and Lloyd C.~L. Hollenberg.
\newblock Scanning {Nanospin} {Ensemble} {Microscope} for {Nanoscale} {Magnetic} and {Thermal} {Imaging}.
\newblock {\em Nano Letters}, 16(1):326--333, January 2016.

\bibitem{petriniNanodiamond2022}
Giulia Petrini, Giulia Tomagra, Ettore Bernardi, Ekaterina Moreva, Paolo Traina, Andrea Marcantoni, Federico Picollo, Klaudia Kvaková, Petr Cígler, Ivo~Pietro Degiovanni, Valentina Carabelli, and Marco Genovese.
\newblock Nanodiamond–{Quantum} {Sensors} {Reveal} {Temperature} {Variation} {Associated} to {Hippocampal} {Neurons} {Firing}.
\newblock {\em Advanced Science}, 9(28):2202014, October 2022.

\bibitem{mommaVESTA2011}
Koichi Momma and Fujio Izumi.
\newblock \textit{{VESTA} 3} for three-dimensional visualization of crystal, volumetric and morphology data.
\newblock {\em Journal of Applied Crystallography}, 44(6):1272--1276, December 2011.

\bibitem{benedicicSpinlattice2025}
Izidor Benedičič, Yuri Tanuma, Žiga Gosar, Bastien Anézo, Mariusz Mrózek, Adam Wojciechowski, and Denis Arčon.
\newblock Spin-lattice relaxation of {NV} centers in nanodiamonds adsorbed on conducting and nonconducting surfaces.
\newblock {\em Physical Review B}, 111(23):235421, June 2025.

\bibitem{bucherQuantum2019}
Dominik~B. Bucher, Diana P.~L. Aude~Craik, Mikael~P. Backlund, Matthew~J. Turner, Oren Ben~Dor, David~R. Glenn, and Ronald~L. Walsworth.
\newblock Quantum diamond spectrometer for nanoscale {NMR} and {ESR} spectroscopy.
\newblock {\em Nature Protocols}, 14(9):2707--2747, September 2019.

\bibitem{mrozekLongitudinal2015}
Mariusz Mrózek, Daniel Rudnicki, Pauli Kehayias, Andrey Jarmola, Dmitry Budker, and Wojciech Gawlik.
\newblock Longitudinal spin relaxation in nitrogen-vacancy ensembles in diamond.
\newblock {\em EPJ Quantum Technology}, 2(1):22, December 2015.

\bibitem{acostaTemperature2010}
V.~M. Acosta, E.~Bauch, M.~P. Ledbetter, A.~Waxman, L.-S. Bouchard, and D.~Budker.
\newblock Temperature {Dependence} of the {Nitrogen}-{Vacancy} {Magnetic} {Resonance} in {Diamond}.
\newblock {\em Physical Review Letters}, 104(7):070801, February 2010.

\bibitem{chenTemperature2011}
X.-D. Chen, C.-H. Dong, F.-W. Sun, C.-L. Zou, J.-M. Cui, Z.-F. Han, and G.-C. Guo.
\newblock Temperature dependent energy level shifts of nitrogen-vacancy centers in diamond.
\newblock {\em Applied Physics Letters}, 99(16):161903, October 2011.

\bibitem{dohertyTemperature2014}
M.~W. Doherty, V.~M. Acosta, A.~Jarmola, M.~S.~J. Barson, N.~B. Manson, D.~Budker, and L.~C.~L. Hollenberg.
\newblock Temperature shifts of the resonances of the {NV} − center in diamond.
\newblock {\em Physical Review B}, 90(4):041201, July 2014.

\bibitem{schafer-nolteTracking2014}
Eike Schäfer-Nolte, Lukas Schlipf, Markus Ternes, Friedemann Reinhard, Klaus Kern, and Jörg Wrachtrup.
\newblock Tracking {Temperature}-{Dependent} {Relaxation} {Times} of {Ferritin} {Nanomagnets} with a {Wideband} {Quantum} {Spectrometer}.
\newblock {\em Physical Review Letters}, 113(21):217204, November 2014.

\bibitem{johnstonStretched2006}
D.~C. Johnston.
\newblock Stretched exponential relaxation arising from a continuous sum of exponential decays.
\newblock {\em Physical Review B}, 74(18):184430, November 2006.

\bibitem{walsh_method_2026}
Ella~P. Walsh, Sepehr Ahmadi, Alexander~J. Healey, David~A. Simpson, and Liam~T. Hall.
\newblock A {Method} {For} {Robust} {Spin} {Relaxometry} in the {Presence} of {Imperfect} {State} {Preparation}.
\newblock {\em Advanced Quantum Technologies}, 9(5):e70291, May 2026.

\bibitem{awadallahSpinstrain2023}
Asad Awadallah, Inbar Zohar, and Amit Finkler.
\newblock Spin-strain coupling in nanodiamonds as a unique cluster identifier.
\newblock {\em Journal of Applied Physics}, 133(14):145103, April 2023.

\bibitem{mittigaImaging2018}
T.~Mittiga, S.~Hsieh, C.~Zu, B.~Kobrin, F.~Machado, P.~Bhattacharyya, N. Z. Rui, A.~Jarmola, S.~Choi, D.~Budker, and N. Y. Yao.
\newblock Imaging the {Local} {Charge} {Environment} of {Nitrogen}-{Vacancy} {Centers} in {Diamond}.
\newblock {\em Physical Review Letters}, 121(24):246402, December 2018.

\bibitem{udvarhelyiSpinstrain2018}
Péter Udvarhelyi, V.~O. Shkolnikov, Adam Gali, Guido Burkard, and András Pályi.
\newblock Spin-strain interaction in nitrogen-vacancy centers in diamond.
\newblock {\em Physical Review B}, 98(7):075201, August 2018.

\bibitem{zhuSimulation2023}
Yuchun Zhu, Elena Losero, Christophe Galland, and Valentin Goblot.
\newblock Simulation of {ODMR} {Spectra} from {Nitrogen}-{Vacancy} {Ensembles} in {Diamond} for {Electric} {Field} {Sensing}, January 2023.
\newblock arXiv:2301.04106 [quant-ph].

\bibitem{thielProbing2019}
L.~Thiel, Z.~Wang, M.~A. Tschudin, D.~Rohner, I.~Gutiérrez-Lezama, N.~Ubrig, M.~Gibertini, E.~Giannini, A.~F. Morpurgo, and P.~Maletinsky.
\newblock Probing magnetism in {2D} materials at the nanoscale with single-spin microscopy.
\newblock {\em Science}, 364(6444):973--976, June 2019.

\bibitem{fincoSingle2023}
Aurore Finco and Vincent Jacques.
\newblock Single spin magnetometry and relaxometry applied to antiferromagnetic materials.
\newblock {\em APL Materials}, 11(10):100901, October 2023.

\bibitem{schmid-lorchRelaxometry2015}
Dominik Schmid-Lorch, Thomas Häberle, Friedemann Reinhard, Andrea Zappe, Michael Slota, Lapo Bogani, Amit Finkler, and Jörg Wrachtrup.
\newblock Relaxometry and {Dephasing} {Imaging} of {Superparamagnetic} {Magnetite} {Nanoparticles} {Using} a {Single} {Qubit}.
\newblock {\em Nano Letters}, 15(8):4942--4947, August 2015.

\bibitem{shenderovaCommercial2017}
Olga Shenderova, Nicholas Nunn, Thomas Oeckinghaus, Marco Torelli, Gary McGuire, Kevin Smith, Evgeny Danilov, Rolf Reuter, Joerg Wrachtrup, Alexander Shames, Daria Filonova, and Alexander Kinev.
\newblock Commercial quantities of ultrasmall fluorescent nanodiamonds containing color centers.
\newblock page 1011803, San Francisco, California, United States, February 2017.

\bibitem{nunnBrilliant2019}
Nicholas Nunn, Neeraj Prabhakar, Philipp Reineck, Valentin Magidson, Erina Kamiya, William~F. Heinz, Marco~D. Torelli, Jessica Rosenholm, Alexander Zaitsev, and Olga Shenderova.
\newblock Brilliant blue, green, yellow, and red fluorescent diamond particles: synthesis, characterization, and multiplex imaging demonstrations.
\newblock {\em Nanoscale}, 11(24):11584--11595, 2019.

\bibitem{jarmolaLongitudinal2015}
A.~Jarmola, A.~Berzins, J.~Smits, K.~Smits, J.~Prikulis, F.~Gahbauer, R.~Ferber, D.~Erts, M.~Auzinsh, and D.~Budker.
\newblock Longitudinal spin-relaxation in nitrogen-vacancy centers in electron irradiated diamond.
\newblock {\em Applied Physics Letters}, 107(24):242403, December 2015.

\bibitem{janiOptically2024}
Mona Jani, Zuzanna Orzechowska, Mariusz Mrózek, Marzena Mitura-Nowak, Wojciech Gawlik, and Adam~M. Wojciechowski.
\newblock Optically detected magnetic resonance study of thermal effects due to absorbing environment around nitrogen-vacancy-nanodiamond powders.
\newblock {\em Journal of Applied Physics}, 135(20):203101, May 2024.

\bibitem{cambriaPhysically2023}
M.~C. Cambria, G.~Thiering, A.~Norambuena, H.~T. Dinani, A.~Gardill, I.~Kemeny, V.~Lordi, A.~Gali, J.~R. Maze, and S.~Kolkowitz.
\newblock A physically motivated analytical expression for the temperature dependence of the zero-field splitting of the nitrogen-vacancy center in diamond, June 2023.
\newblock arXiv:2306.05318 [cond-mat].

\bibitem{jegenyesMaterials2026}
Nikoletta Jegenyes, Vladimir Verkhovlyuk, Szabolcs Czene, Attila Csáki, Olga Krafcsik, Zsolt Czigány, David Beke, and Adam Gali.
\newblock Materials and {Spin} {Characteristics} of {Nanodiamonds} {Partially} {Covered} with {Amino} {Groups} and {Embedded} with {Nitrogen}-{Vacancy} {Color} {Centers}.
\newblock {\em ACS Omega}, 11(35):53440--53451, September 2026.
\newblock arXiv:2502.07706 [quant-ph].

\bibitem{acostaOptical2010}
V.~M. Acosta, A.~Jarmola, E.~Bauch, and D.~Budker.
\newblock Optical properties of the nitrogen-vacancy singlet levels in diamond.
\newblock {\em Physical Review B}, 82(20):201202, November 2010.

\bibitem{dohertyNitrogenvacancy2013}
Marcus~W. Doherty, Neil~B. Manson, Paul Delaney, Fedor Jelezko, Jörg Wrachtrup, and Lloyd~C.L. Hollenberg.
\newblock The nitrogen-vacancy colour centre in diamond.
\newblock {\em Physics Reports}, 528(1):1--45, July 2013.

\bibitem{vanoortElectricfieldinduced1990}
Eric Van~Oort and Max Glasbeek.
\newblock Electric-field-induced modulation of spin echoes of {N}-{V} centers in diamond.
\newblock {\em Chemical Physics Letters}, 168(6):529--532, May 1990.

\bibitem{feltonHyperfine2009}
S.~Felton, A.~M. Edmonds, M.~E. Newton, P.~M. Martineau, D.~Fisher, D.~J. Twitchen, and J.~M. Baker.
\newblock Hyperfine interaction in the ground state of the negatively charged nitrogen vacancy center in diamond.
\newblock {\em Physical Review B}, 79(7):075203, February 2009.

\bibitem{tsukamotoAccurate2022}
Moeta Tsukamoto, Shuji Ito, Kensuke Ogawa, Yuto Ashida, Kento Sasaki, and Kensuke Kobayashi.
\newblock Accurate magnetic field imaging using nanodiamond quantum sensors enhanced by machine learning.
\newblock {\em Scientific Reports}, 12(1):13942, September 2022.

\end{thebibliography}

\clearpage
\appendix  
\renewcommand{\thefigure}{S\arabic{figure}}
\renewcommand{\thetable}{S\arabic{table}}
\renewcommand{\theequation}{S\arabic{equation}}
\setcounter{figure}{0}
\setcounter{table}{0}
\setcounter{equation}{0}

\begin{center}
  {\Large\bfseries Supporting Information}\\[1.2em]
  {\large\bfseries Quantum sensing of weak ferromagnetism in a phase-separated Mn-doped \ce{CaFe3O5}}\\[1em]
  Izidor Benedičič,$^{1}$, J. Paul Attfield$^{2}$, Denis Arčon$^{1,3}$\\[0.5em]
  {\small $^{1}$Institut Jožef Stefan, Jamova c. 39, 1000 Ljubljana, Slovenia\\
  $^{2}$Centre for Science at Extreme conditions and School of Chemistry, University of Edinburgh, Mayfield Road, Edinburgh EH9 3FD, United  Kingdom\\
  $^{3}$Fakulteta za matematiko in fiziko, Jadranska c 19, 1000 Ljubljana, Slovenia}
\end{center}
\vspace{1em}
\hrule
\vspace{1.5em}



\section{Optical and electron microscopy images of the sample}

\begin{figure}[H]
	\centering
	\includegraphics[width=0.8\textwidth]{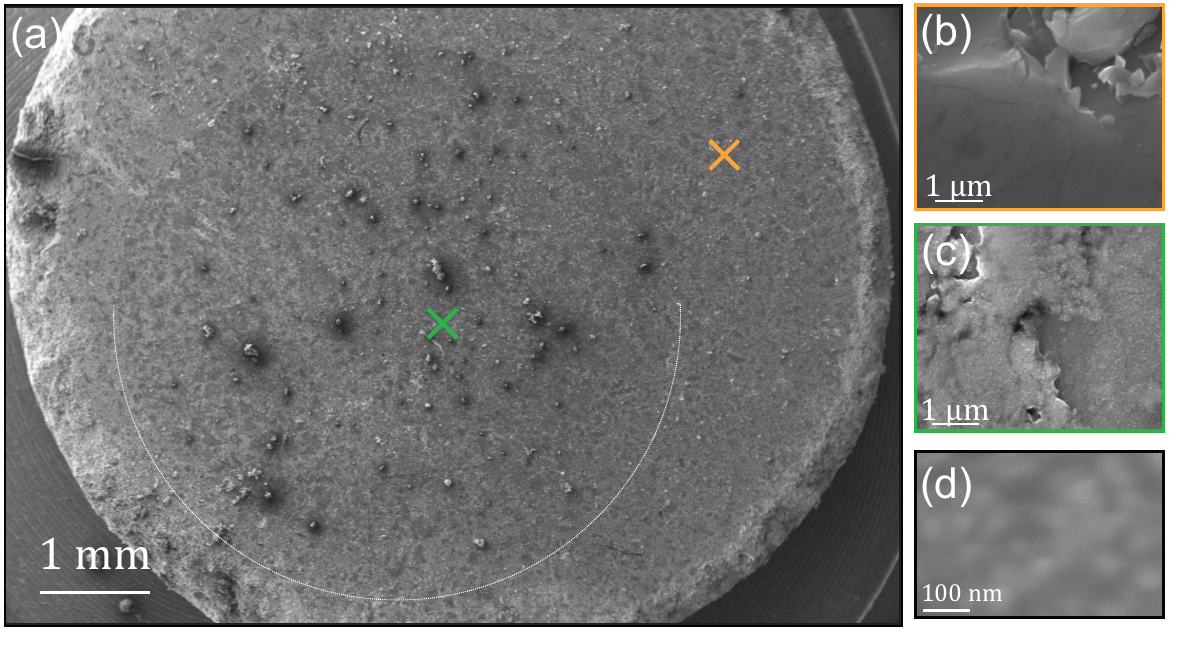}  
	\caption{ (a) Scanning electron microscopy image of the sample. The nanodiamonds were deposited on the darker circular area, marked with a white semicircle for orientation. Additional close-up images were taken of the areas denoted with green and yellow crosses. Large particles can be seen on the surface, but they are well separated. Their large size indicates they are not nanodiamond aggregates.
    (b) Closeup of a $\mathunit{\mu m}$-scale area where no diamonds were deposited.
    (c) Closeup of a $\mathunit{\mu m}$-scale area within the region with deposited nanodiamonds. Nanodiamond particles can be seen as brighter spots on the darker background. No large clusters can be observed on these length scales.
    (d) Closeup of a $100\mathunit{nm}$-scale area, where individual nanodiamond particles can be observed as bright flecks.
    }
	\label{fig:semImages}
\end{figure}

The Mn-doped \ce{CaFe3O5} pellet with indented nanodiamonds was investigated under a scanning electron microscope (SEM) to check for the presence of nanodiamond clustering. On larger scales, such as in Figure \ref{fig:semImages} (a), several well-separated pieces of debris can be visible. Their sizes and distributions suggest these are not agglomerations of nanodiamonds but rather crumbled-off pieces of the powder pellet. A close-up image of the area without deposited nanodiamonds is shown in Figure \ref{fig:semImages} (b) for reference, and an image of a nanodiamond-covered area on the same scale is shown in Figure \ref{fig:semImages} (c). The nanodiamonds uniformly cover the surface, with occasional patches with a smaller surface density. Under the maximum magnification of our SEM instrument, Figure \ref{fig:semImages} (d), individual nanodiamond particles with a diameter of about $40 \mathunit{nm}$ can be clearly observed. These SEM images thus exclude any nanodiamond clustering on the surface of Mn-doped \ce{CaFe3O5}.

\pagebreak
\section{Measured variations across the sample}

\begin{figure}[H]
	\centering
	\includegraphics[width=0.67\textwidth]{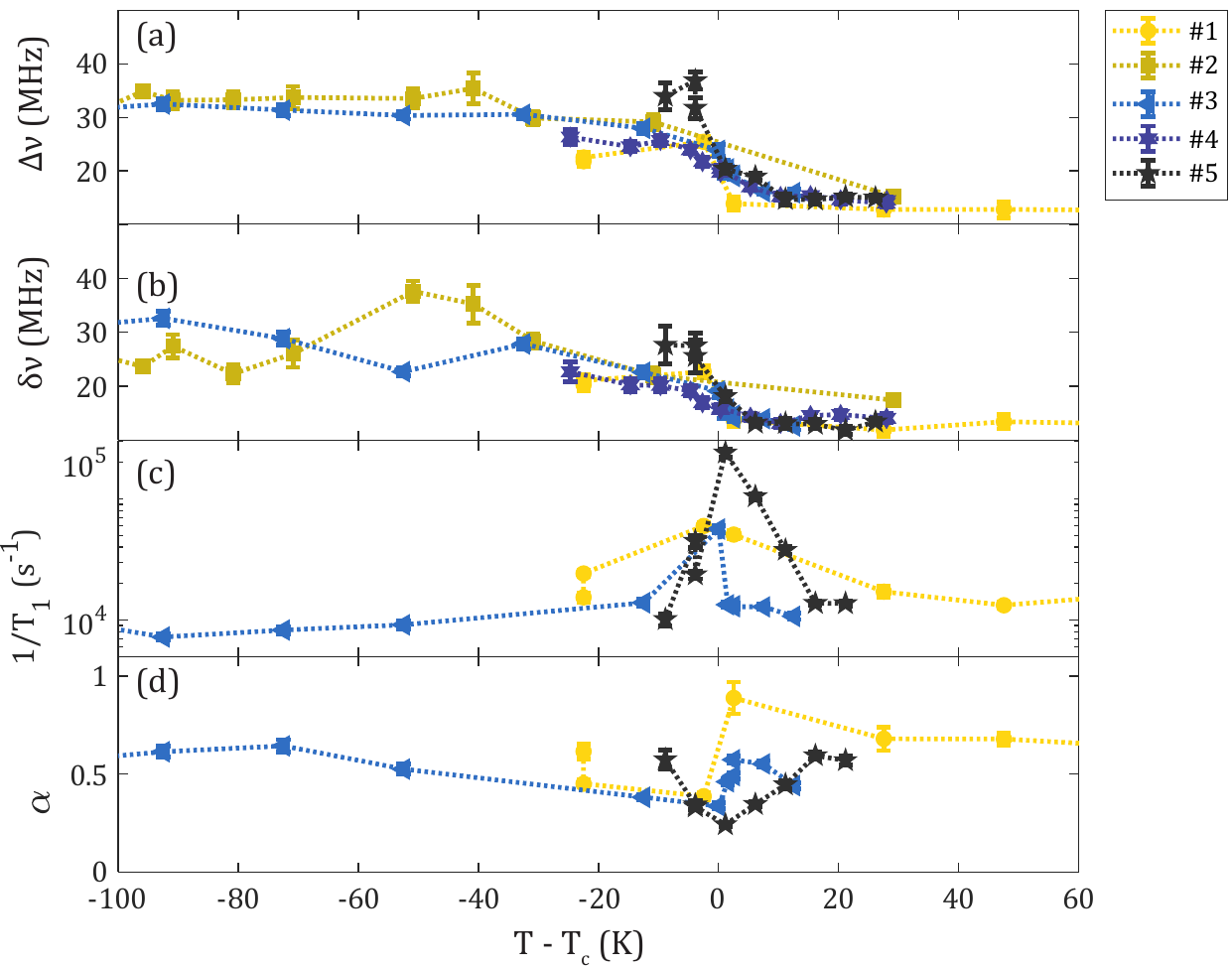}  
	\caption{ (a) Temperature dependence of the splitting of the two NV-ODMR spectra, $\Delta\nu = \nu_+-\nu_-$, obtained after fitting the spectra with two overlapping Lorentzian curves.
    (b) Temperature dependence of the full width at half maximum (FWHM) $\delta\nu$ of the Lorentzian curves fitted to the two peaks in the NV-ODMR spectra. For clarity, only the width of the peak at $\nu_+$ is plotted.
    (c) Temperature dependence of the spin-lattice relaxation rates $1/T_1 $ obtained from fitting of NV-magnetization curves with a stretched exponential curve. The sharp enhancement of $1/T_1$ at $T=T_\mathrm{c}$ is consistent with the divergence of the dynamic spin susceptibility due to the critical slowing down of spin dynamics.
    (d) Temperature dependence of the stretching exponent $\alpha$ obtained from fitting of NV-magnetization curves to the stretched relaxation model (see main text for details). The dip in the temperature dependence of $\alpha$ at $T=T_\mathrm{c}$ points towards a broader distribution of relaxation rates as the ferromagnetic transition in the CA phase sets in.
    }
	\label{fig:spotComparisons}
\end{figure}

Qualitatively, the same temperature dependence of the optically detected magnetic resonance (ODMR) spectra is observed at multiple locations on the sample; however, the precise transition temperature and the magnitude of additional broadening, discussed in the main text (Figure 2), vary. To compare the data collected for different measurement locations, we plot the splitting in the NV-ODMR spectra $\Delta\nu(T)$ as a function of temperature and fit it to an empirical arctangent function 
\begin{equation}
    \Delta\nu(T) = \Delta\nu_\infty + A\arctan\left(\frac{T-T_\mathrm{c}}{\Delta T}\right), \label{eq:transitionFit}
\end{equation}
where $1/\Delta T$ is the scaling factor describing the transition width and $T_\mathrm{c}$ is the transition temperature. The two parameters $\Delta\nu_\infty$ and $A$ relate to the limiting values of $\Delta\nu(T)$ in the paramagnetic and deep in the ferromagnetic phase. The values of $T_\mathrm{c}$ obtained from such a fit are collected in Table \ref{table:tcvalues}.  When plotted on the common temperature scale $T-T_\mathrm{c}$ (Figure \ref{fig:spotComparisons}), we observe a consistent change of spectra across the magnetic transition, although with some variations in magnitudes of both splitting and broadening. 

\section{Local heating effects}

The values of $T_\mathrm{c}$ obtained from an empirical fit to Eq. \ref{eq:transitionFit} range from $T = 223 - 273 ~\mathrm{K}$. This variation in the apparent transition temperature can be explained by the experimental conditions. The temperature probe in the cryostat is mounted at the bottom of the cold finger, on which the sample is positioned. The cold finger has high thermal conductivity, and a thermally conductive vacuum grease was used to ensure good thermal contact with the sample. However, the used pellet itself is still relatively thick ($d = 1.6 \mathunit{mm}$). Moreover, the studied material is a poor conductor ($\rho \sim 10^{-3} \mathunit{\Omega cm}$ at $T=300\mathunit{K}$), thus the electronic contribution to thermal conductivity is small. The multi-grain structure of the pellet additionally hinders efficient thermal coupling between the cold finger and the top of the sample, where the nanodiamonds are embedded.\par
At the same time, the measured spot is subject to local heating due to laser illumination and absorption of microwaves; thereby, the sample temperature may locally increase relative to the nominal temperature of the temperature sensor. A similar effect was recently observed by Jani \textit{et al.} \cite{janiOptically2024} in a system of nanodiamonds mixed with \ce{Fe2O3} or graphite powder: in their wide-field setup, they report temperature shifts of between $-1.73$ and $-2.26 \mathunit{K/mW}$, depending on the environment.\par
\begin{figure}[H]
	\centering
	\includegraphics[width=\textwidth]{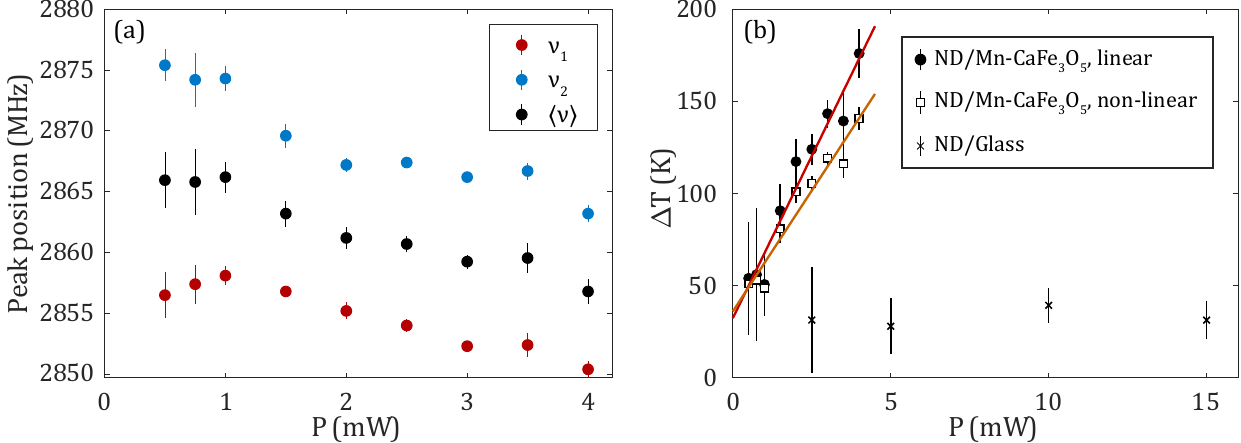}  
	\caption{ (a) Laser power dependence of the peaks fitted to the NV-ODMR spectrum. Error bars represent the fitting uncertainty.
    (b) Temperature shift as a function of laser power, calculated from data in panel (a). Full circles were calcualted using values from Acosta \textit{et al.} ($dD/dT = - 75 \mathunit{kHz/K}$, $D_{(T=300 \mathunit{K})} = 2870 \mathunit{MHz} $)\cite{acostaTemperature2010}. At large $\Delta T$, linear dependence no longer holds, and we include a calculation using the expression from Ref. \cite{cambriaPhysically2023} (empty squares). The slopes of linear fits (red and orange lines) are approximately $\approx 35 \mathunit{K/mW}$ and $\approx 28 \mathunit{K/mW}$, respectively. Black crosses denote measurements of nanodiamonds from the same batch deposited on glass.
    }
	\label{fig:laserHeating}
\end{figure}

In our confocal setup, the average energy flux density is expected to be significantly higher. We estimate this effect by measuring the ODMR spectra using varying laser power. Figure \ref{fig:laserHeating} (a) shows the fitted peak positions as a function of temperature, and a trend of decreasing peak frequency with increased power is clearly observed. At lower power ($P<1 \mathunit{mW}$), the effect seems to saturate, though the quality of fit decreases due to a lower overall signal level. At the same time, additional splitting is observed, which could originate from closer proximity to the magnetic transition at $T_m = 302 \mathunit{K}$. The relative temperature shift, calculated using values from Ref. \cite{cambriaPhysically2023}, is plotted in Fig. \ref{fig:laserHeating} (b), showing a slope of $\approx 26.2 \mathunit{K/mW}$, and extrapolates to $\Delta T = 36.1 \mathunit{K}$ at $P = 0 $. Reference measurements of the same batch of nanodiamonds deposited on a glass substrate (cross symbols in Fig. \ref{fig:laserHeating} (b)) show the central frequency is essentially power-independent, in agreement with findings from Jani \textit{et al.} \cite{janiOptically2024}.
This suggests the apparent frequency shift at $P = 0$ is not an artifact of heating but should rather be taken as a reference value for the zero-field splitting for our particular batch of nanodiamonds: $D_{(T = 293 \mathunit{K})} = 2867.6 \mathunit{MHz}$. Such discrepancies from the bulk value are in fact commonly found in nanodiamonds \cite{awadallahSpinstrain2023,jegenyesMaterials2026}.

\begin{center}
\begin{table}[h]
\[
\begin{array}{|c||c|c|c|c|c|c|} 
    \hline
    \text{Location} & 1 & 2 & 3 & 4 & 5  \\
    \hline
    T_{\rm c} ~(\rm K) & 223 & 271 & 273 & 265 & 244   \\ 
    \hline 
\end{array}
\]
\caption{Measured nominal magnetic transition temperatures}
\label{table:tcvalues}
\end{table}
\end{center}

\begin{figure}[H]
	\centering
	\includegraphics[width=\textwidth]{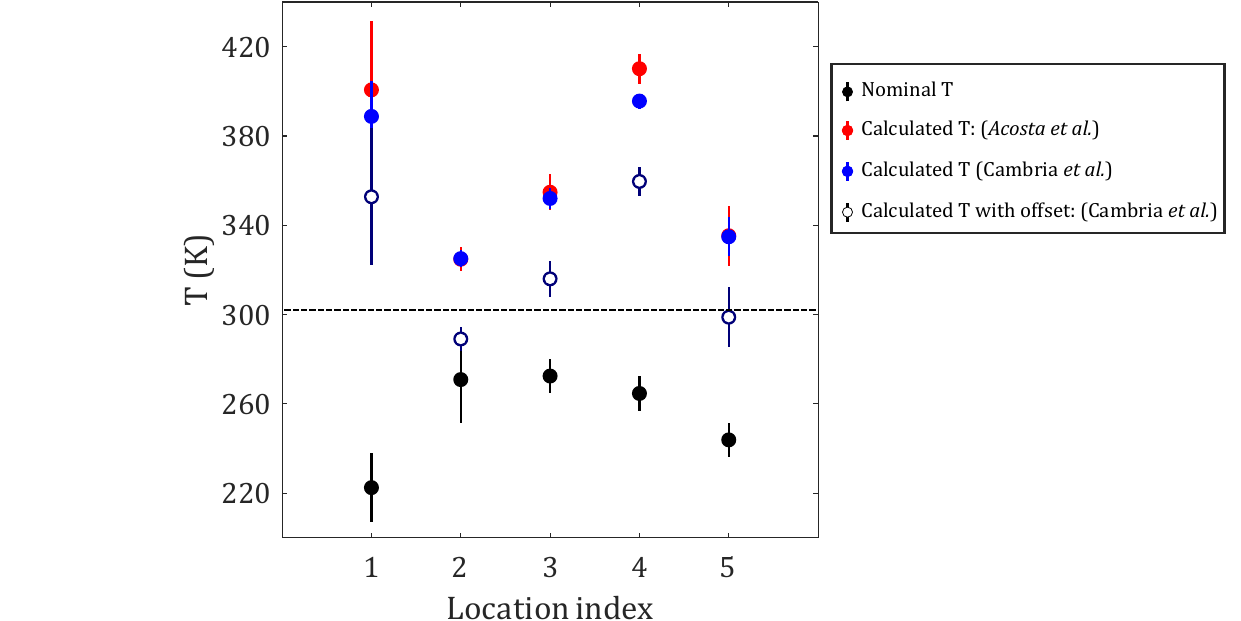}  
	\caption{ Ferromagnetic transition temperatures determined from nominal temperature reading (black symbols) and calculated from the spectral frequency shift (red and blue symbols). The calculated temperature was obtained by fitting two Lorentzian peaks to the spectrum taken at the temperature closest to the fitted $T_c$. The mean frequency $\langle \nu \rangle$ was then inserted in the formula from Refs. \cite{acostaTemperature2010} (red) and \cite{cambriaPhysically2023} (blue). The empty blue circles were obtained by taking into account that the low-power room-temperature central frequency is lower than the literature value of $D=2.870 \mathunit{GHz}$. The dashed line denotes the transition temperature for weak ferromagnetism from bulk measurements. 
    }
	\label{fig:transitionT}
\end{figure}

In Fig. \ref{fig:transitionT}, we also show the discrepancy between the measured nominal transition temperature (black symbols) and the temperature calculated from the central frequency shift (red and blue symbols). The calculated values were obtained using formulas from Refs. \cite{acostaOptical2010} (red) and \cite{cambriaPhysically2023} (blue). The nominal temperatures, measured at the bottom of the cold finger, are always lower than the inferred values, supporting the hypothesis of a thermal gradient. However, the temperature offset is not the same at all measurement locations, suggesting an additional effect, possibly local variations in intrinsic nanodiamond strain.\par
By accounting for the fact that our nanodiamond sample has an apparently different $D_0$, as shown in Fig. \ref{fig:laserHeating} (b), we can further subtract this offset of $\Delta T = 36 \mathunit{K}$, resulting in values plotted as dark blue empty circles. Such final estimation brings the local temperature at three of the locations within $\delta T = 10 \mathunit{K}$ of the expected $T_m$.
However, two measurement points still show a discrepancy of $T_\mathrm{corrected}-T_m \approx 60 \mathunit{K}$, corresponding to a frequency shift of $-4.5 \mathunit{MHz}$, suggesting another effect is still unaccounted for.

\pagebreak
\section{Magnetic effects of compression and thermal cycling}

\begin{figure}[H]
	\centering
	\includegraphics[width=0.7\textwidth]{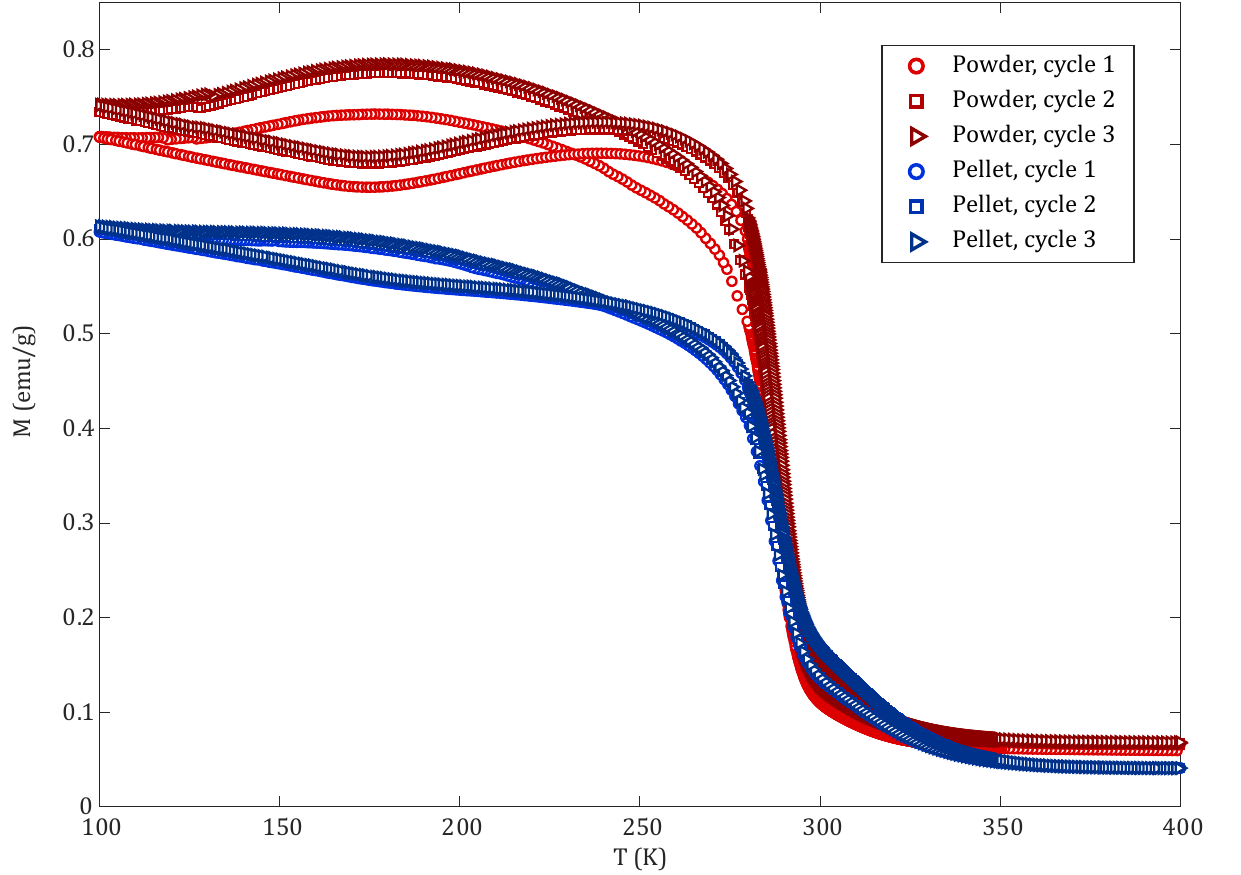}  
	\caption{Curves of magnetization as a function of temperature measured on the powder sample (red symbols) and on the pellet after compression (blue). For each sample, we measured three temperature cycles, each consisting of field-cooling ($H=  100 \mathunit{Oe}$) from 400 to 100 K and heating back to 400 K. The  Thermal cycles have only slight effects on the total measured magnetization. Total magnetization in the pellet sample is $\approx 16\%$ lower than in the powder sample.
    }
	\label{fig:magnetization}
\end{figure}

To control for the effects of thermal cycling and increased pressure during the sample preparation, we measured the magnetization as a function of temperature of the powder sample and for a piece of the pellet after compression. To minimize the signal from paramagnetic impurities, we performed the measurements on a piece from the edge of the pellet, were no nanodiamonds were deposited. The $M(T)$ curves are presented in Fig. \ref{fig:magnetization}: while the weak ferromagnetic transition is observed in both samples, the powder sample (plotted with red symbols) shows higher saturation magnetization as well as a sharper transition. The decrease in magnetization could be a consequence of spin-strain coupling changing the balance of the electronic phases during the sample preparation.\par
For each sample, we measured three $M(T)$ in three temperature cycles, each consisting of field-cooling ($H=  100 \mathunit{Oe}$) from 400 to 100 K and heating back to 400 K. Close to the transition, we observe small hysteresis effects on the order of a few $\mathrm{K}$. These effects are too small to explain the locally observed variations in nominal transition temperature, which are therefore more likely to be dependent on local thermal conductivity and absorption properties. Furthermore, the magnetization magnitude changes very little in the pellet sample, corroborating that the observed variations in spectral widths originate from local differences instead of average properties of the sample.

\section{Fitting of spin-lattice relaxation curves}

\begin{figure}[H]
	\centering
	\includegraphics[width=\textwidth]{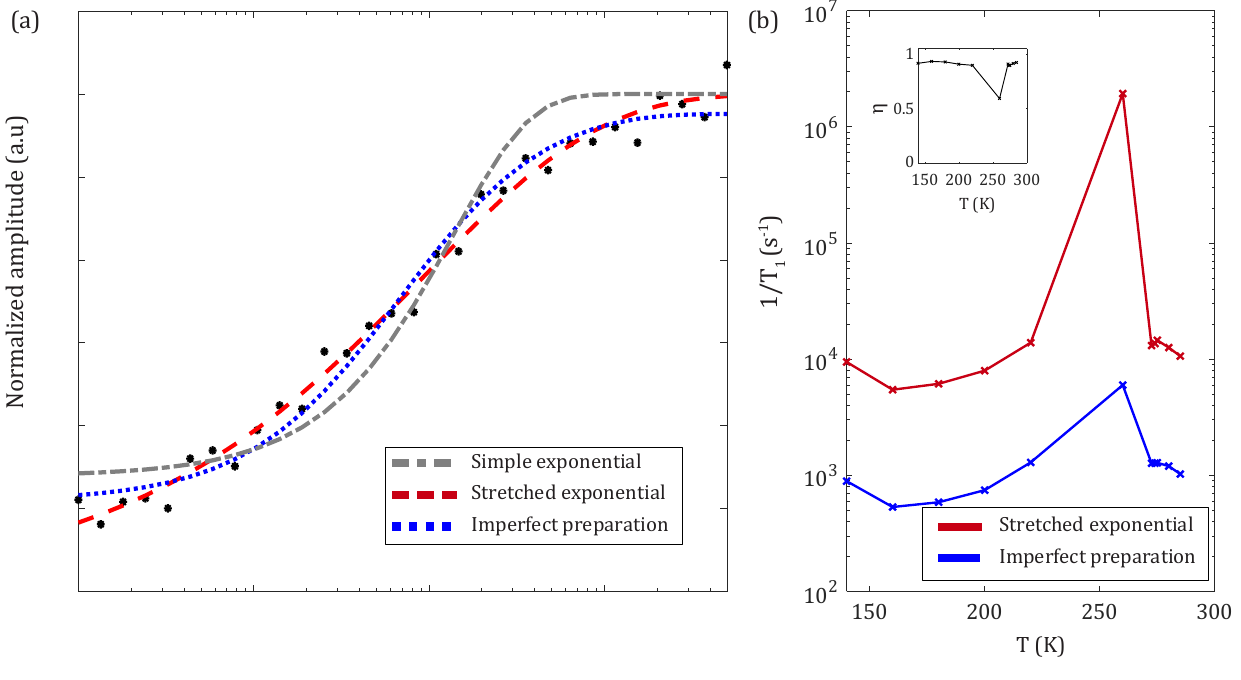}  
	\caption{(a) Comparison of the three relaxation models fitted to an experimental relaxation curve. The stretched-exponential curve (red) yields the lowest residual sum of squares, followed by the imperfect preparation model (blue). The single exponential model (gray) performs the worst.
    (b) Spin-lattice relaxation rate $1/T_1$ from stretched-exponential and imperfect relaxation model. Both curves have qualitatively the same behavior but differ in the value for $1/T_1$. Insert: temperature dependence of the imperfection parameter $\eta$.
    }
	\label{fig:curveFits}
\end{figure}

To find the most appropriate fitting curve for the data acquired during the relaxometry measurement, we fitted three candidate functions (Figure \ref{fig:curveFits} (a): a single-exponential curve of the form $A= A_0e^{-\tau/T_1}$, the stretched exponential curve $A=A_0 e^{-(Cen\tau/T_1)^\alpha}$, where $0 < \alpha < 1$, and the relaxation curve based on an imperfect state preparation model in Ref. \cite{walsh_method_2026}, $A = \frac{\eta-1}{\eta - B \exp(\tau/T_1)}$, where $\eta$ is the imperfection parameter. The single-exponential curve (gray dashed line) notably deviates from the data point in the region between $\tau = 100 ~\mathrm{\mu s}$ and $\tau = 1 ~\mathrm{ms}$. Both the stretched exponential (red dashed) and the imperfect preparation (blue dotted) curves fit the data well over the entire range of $\tau$. \par
By comparing the sum of squared errors (SSE), we find the stretched exponential curve (SSE = 0.04) fits the data better than the imperfect preparation curve (SSE = 0.07), and both perform better than the single-exponent fit (SSE = 0.20). In the main text, we report the values of $T_1$ from the stretched exponential fit. However, we note that the results would not be qualitatively different if we choose the imperfect preparation model instead, as is illustrated in Fig. \ref{fig:curveFits} (b): though numerical values are different, the relaxation rates show the same enhancement at the magnetic transition.

\section{Simulations of the ODMR spectra}

\subsection{Microscopic Hamiltonian}

The ODMR spectra were calculated by diagonalizing the full Hamiltonian of the NV center. The ground-state spin Hamiltonian of the NV center  is
\begin{equation}
    H = H_\mathrm{el} + H_\mathrm{nuc},
\end{equation}
where $H_\mathrm{el}$ is the electronic spin Hamiltonian and $ H_\mathrm{nuc} $ is the hyperfine interaction between the electronic spin of the NV centre and the nitrogen nuclear spin. In the presence of an external magnetic field $\Vec{B}$, electric field $\Vec{E}$ and strain tensor $\mathbf{\varepsilon}$, the electronic spin Hamiltonian can be written as \cite{dohertyNitrogenvacancy2013}
\begin{equation}\label{eq:hamiltonian}
    \frac{1}{h}H_\mathrm{el} = (D_\mathrm{gs} + d_{||}E_z) \left( S_z^2 - \frac{2}{3} \right)  + \gamma_\mathrm{NV} \Vec {B} \cdot \Vec {S} - d_\perp E_x(S_x^2 - S_y^2) + d_\perp E_y(S_xS_y + S_yS_x) + \frac{1}{h}H_\mathrm{strain},
\end{equation}
where $D_\mathrm{gs}$ is the temperature-dependent zero-field splitting, $\gamma_\mathrm{NV} = 28 \mathunit{GHz/T}$ is the gyromagnetic ratio, $h$ is the Planck constant, and $d_{||}$ and  $d_{\perp}$ are the electric coupling constants for parallel and transverse components. The NV center has a total spin of $S=1$ and the electronic spin operators $S_x, S_y$, and $S_z$ can be represented in a standard matrix form:
\begin{equation}
S_x= \frac{1}{\sqrt{2}}
    \begin{pmatrix}
    0 & 1 & 0\\
    1 & 0 & 1\\
    0 & 1 & 0
    \end{pmatrix} ,
Sy = \frac{1}{\sqrt{2}}
    \begin{pmatrix}
    0 & -i & 0\\
    i & 0 & -i\\
    0 & i & 0
    \end{pmatrix} 	,
S_z = 
    \begin{pmatrix}
    1 & 0 & 0\\
    0 & 0 & 0\\
    0 & 0 & -1
    \end{pmatrix} 	.
\end{equation}
The spin-strain coupling Hamiltonian is  

\begin{equation} \label{eq:spinstrain}
\begin{split}
H_\mathrm{strain} = & [ h_{41}(\epsilon_{xx} + \epsilon_{yy}) + h_{43}\epsilon_{zz}]S_z^2 \\
&+ \tfrac{1}{2}[ h_{26}\epsilon_{xz} - \tfrac{1}{2}h_{25}(\epsilon_{xx} - \epsilon_{yy}) ] \{ S_x, S_z \} + \tfrac{1}{2}(h_{26}\epsilon_{yz} +  h_{25}\epsilon_{xy})\{ S_y, S_z \} \\
&+ \tfrac{1}{2}[ h_{16}\epsilon_{xz} - \tfrac{1}{2}h_{15}(\epsilon_{xx} - \epsilon_{yy}) ] \left( S_y^2 - S_x^2 \right) + \tfrac{1}{2}(h_{16}\epsilon_{yz} +  h_{15}\epsilon_{xy})\{ S_x, S_y \},
\end{split}
\end{equation}
where $\epsilon_{ij}$ are the strain tensor components and $h_{ij}$ the corresponding coupling constants. Lattice strain can, in principle, be accounted for as an effective electric field, because the spin-strain and spin-electric Hamiltonian are symmetry-equivalent. The response to the local electric field is strongly anisotropic ($d_\perp/d_\parallel \approx 50$), whereas anisotropies in strain coupling are less pronounced. We thus included each part of the Hamiltonian separately. However, we note that the ambiguity remains - different combinations of strain and electric field distributions could result in the same final absorption spectrum. This does not affect our main findings, as the aim of the work is not to determine internal strain and electric fields within nanodiamonds, but to reconstruct the magnetic fields. It is thus sufficient to obtain parameters that satisfactorily describe the high-temperature spectra that can then be further modified by adding the magnetic field. \par 
Finally, the \ce{^{14}N} nuclear Hamiltonian is 
\begin{equation} \label{eq:hyperfine}
    \frac{1}{h}H_\mathrm{nuc} = A_{||}S_zI_z + A_\perp (S_xI_x + S_yI_y ) + P_Q\left( I_z^2 - \frac{2}{3} \right) +\gamma_N\vec{B}\cdot\vec{I},
\end{equation}
where $A_{||}$ and $A_\perp$ are the \ce{^{14}N} hyperfine parameters, $P_Q$ is the nuclear quadrupole parameter and $\gamma_N = 3.1 \mathunit{MHz/T}$ is the nuclear gyromagnetic ratio.
For completeness, we also included the hyperfine effects, although they are smaller than the relevant electronic effects and could also be neglected without compromising the fit quality. The values of all coupling constants used in the simulation of the ODMR spectra reported here are given in Table \ref{table:coupling}.\par

To obtain the microwave transitions, we additionally introduce a linearly polarized microwave field induced by the MW antenna in the form of an interaction Hamiltonian:
\begin{equation}
    \frac{1}{h}H_\mathrm{int} = \gamma_\mathrm{NV}\vec{B}_\mathrm{MW}\cdot \vec{S} + \gamma_\mathrm{N}\vec{B}_\mathrm{MW} \cdot \vec{I}.
\end{equation}
This interaction Hamiltonian couples the initial states $\ket{i}$ with the final states $\ket{f}$ of the electronic spn Hamiltonian. Namely:
\begin{equation}
    T_{i,f} \propto \left| \bra{f} H_\mathrm{int} \ket{i}\right|^2.
\end{equation}
The transitions were furthermore assumed to have a Lorentzian shape with linewidth $\Gamma$:
\begin{equation}
    T_{i,f}(\nu) = \frac{T_{i,f}\Gamma^2}{\left( (\nu - |\nu_f - \nu_i|)^2 + \Gamma^2 \right)},
\end{equation}
where $T_{i,f}$ is the transition matrix element for the transition between the initial and final states and $\nu_{i,f}$ are their respective frequencies.
The broadening parameter $\Gamma$ depends on the specific experimental conditions; here, it was taken to be $\Gamma  = 1 \mathunit{MHz}$ throughout. The total transition strength was then determined by summing over all possible initial and final states:
\begin{equation}
    T(\nu) = \sum_{i,f}T_{i,f}(\nu).
\end{equation}
We assume all matrix elements are equal, $T_{i,f} = T_0$, determined by fitting to the experimental data. 

\begin{center}
\begin{table}[h]
\[
\begin{array}{|c||c|c|c|c|c|c|} 
 \hline
 \text{Strain coupling constants}  & h_{41} & h_{43} & h_{25} & h_{26} & h_{16} & h_{15} \\ [0.5ex] 
 \hline
 \text{Value} \mathunit{(GHz)} & -6.42 & 2.3  & -2.6  & -2.83 & 5.7  & 19.66 \\ 
 \hline \hline
 \text{Electric coupling constants} & d_{||} & d_\perp & & & & \\
 \hline
 \text{Value} \mathunit{(Hz ~m/V)} & 3.5\cdot10^{-3} & 0.17 & & & &  \\
 \hline \hline
 \text{Hyperfine coupling constants} & A_{||} & A_\perp & P_Q & & & \\
 \hline
 \text{Value} \mathunit{(MHz)} & -2.14 & -2.7 & -4.95 & & & \\
 \hline
\end{array}
\]
\caption{Coupling constants used in Equations \ref{eq:hamiltonian}, \ref{eq:spinstrain} and \ref{eq:hyperfine}. The electric and hyperfine coupling constants were determined experimentally; we use the values reported in \cite{vanoortElectricfieldinduced1990, dohertyNitrogenvacancy2013} and \cite{feltonHyperfine2009}, respectively. The hyperfine coupling constants are reported for the \ce{^{14}N} isotope. The spin-strain coupling constants are reproduced from Ref. \cite{udvarhelyiSpinstrain2018} and were calculated using density functional theory. }
\label{table:coupling}
\end{table}
\end{center}

\subsection{Ensemble averaging}

Within each nanodiamond, there are four possible orientations of NV center axes. In nanodiamonds with nominal diameter $d = 40 \mathunit{nm}$ and concentration of $c = 1.5 \mathunit{ppm}$, there are on average 14 NV centers in each diamond. The diameter of the illuminated spot is also much larger than the diameter of a single nanodiamond; therefore, a large number ($N \approx 10^5$) of diamonds are illuminated simultaneously. To obtain an ensemble average, we generate a random magnetic field vector, electric field vector, and strain tensor for each nanodiamond, and average the transition strength over $N_\mathrm{defect} = 14$ NV centers, each with a direction randomly chosen from the four possible orientations. Finally, we average the transition strengths over a large number of nanodiamonds (N = 2000). The magnitudes of the magnetic field, the electric field, and strain were sampled from normal distributions:
\begin{equation}
     P_A = \frac{1}{\sigma_A \sqrt{2\pi}} \exp \left( -\frac{(A-\mu_{A})^2}{2\sigma_A^2} \right),
\end{equation}
where $P_A$ is the probability density, $\mu_A$ is the mean value, and $\sigma_A$ is the standard deviation of the distribution of quantity $A$, which is either magnetic field density, electric field magnitude, or strain.
The best fit of the high-temperature data was achieved with the combination of parameters $\mu_E = 50 \mathunit{MV/m}$, $\sigma_E = 10 \mathunit{MV/m}$, $\mu_\mathrm{\epsilon} = 0 $ and $\sigma_\epsilon = 5.4\cdot 10^{-3}$, with each component of the strain tensor $\epsilon_{ij}$ drawn independently from the same distribution. The same values for electric field and strain parameters were used in the low-temperature phase but with an added magnetic field. There, we observe notable variations between the measured locations: this is discussed in detail in the main text. The direction vectors for both $\vec{B}$ and $\vec{E}$ were sampled uniformly on the unit sphere. \par

For completeness, we note that we omitted three geometric factors that would be relevant when investigating single nanodiamonds or smaller ensembles. First, the optical transition probability depends on the angle between the excitation electric field and the NV center axis. The transition probability is  proportional to the inner product between the electric field and the electric dipole moment of the NV center, leading to
\begin{equation}
    \kappa(\theta) \propto 1+\cos^2(\theta),
\end{equation}
where $\theta$ is the angle between the NV center axis and the optical axis. Thus, the absorption efficiency is higher when the NV axis is parallel to the optical axis. The second geometric factor originates in the fact that the dipole emission is not isotropic, and the light is more likely to be emitted in the direction perpendicular to the NV axis. In an experiment, light-collection efficiency is limited by the objective lens's numerical aperture, $\mathrm{NA} = \sin(\theta_\mathrm{max})$. By integrating over all observable collection angles, the light collection efficiency can be written as \cite{tsukamotoAccurate2022}
\begin{equation}
    P(\theta) \propto 32 - \left( 31 + \cos(2\theta_\mathrm{max})\cos\theta_\mathrm{max} \right) - 6\cos(2\theta)\sin^2(\theta_\mathrm{max}).
\end{equation}

The first factor favours the signal from NV centers with their axis parallel to the excitation beam, while the second factor suppresses the light emitted from those centers. The total effect of these two factors thus mainly depends on the numerical aperture used in the experiment. In our setup, $\mathrm{NA} = 0.45$, and the product $P(\theta) \cdot \kappa(\theta)$ varies by up to 12 \%. These two geometric factors are therefore a relatively small correction and do not qualitatively change the final results.\par
The third geometric factor that could be included is the magnetic-field polarization of the microwaves driving the transitions. The probability for a magnetic dipole transition between an initial state $\ket{i}$ and a final state $\ket{f}$ is 
\begin{equation}
    T_{i,f} \propto \left| \braket{f|H_\mathrm{int}|i} \right|^2,
\end{equation}
where $H_{int}$ is the interaction Hamiltonian of the form
\begin{equation}
    \frac{1}{h}H_{int} = \gamma_\mathrm{NV}\vec{S}\cdot\vec{B}_\mathrm{MW} + \gamma_\mathrm{N}\vec{I}\cdot\vec{B}_\mathrm{MW},
\end{equation}
where $\vec{B}_\mathrm{MW}$ is the microwave magnetic field. The omega-shaped microwave antenna \cite{benedicicSpinlattice2025} used in the experiment produces a linearly polarized field. However, due to the random orientation of nanodiamonds with respect to the microwave antenna axis, we average over a large number of uniformly distributed unit vectors, and this effect is averaged out.

\section{Two-component fitting}

To fit the ODMR spectrum using two spectral components, we first obtained the two trial spectra. We name them $L(\nu)_\mathrm{AFM}$ and $L(\nu)_\mathrm{WFM}$ to represent the spectra originating from a strictly antiferromagnetic and weakly ferromagnetic phase, respectively. The antiferromagnetic component, $L(\nu)_\mathrm{AFM}$, was obtained by fitting the electric field and strain parameters to high-temperature spectra, as shown in the main text. To obtain the component $L(\nu)_\mathrm{WFM}$, we first chose the mean of the magnetic field distribution as the maximum value fitted across the measurement locations (Location 2 in Fig. 4 (b) in the main text, $B=1.2 \mathunit{mT}$). This choice follows from the model assumptions: when measured entirely within the weakly ferromagnetic domain, we expect a maximal mean magnetic field, and any admixture of NV centers measured on antiferromagnetic domains would decrease it.\par
\begin{figure}[H]
	\centering
	\includegraphics[width=\textwidth]{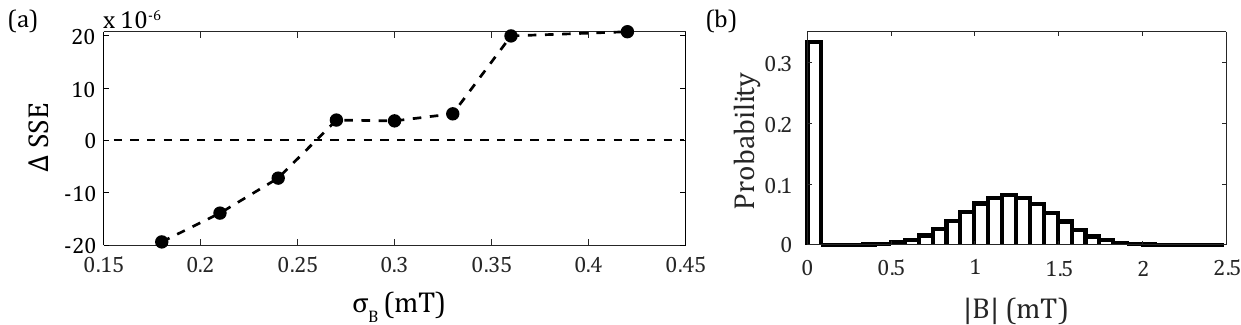}  
	\caption{(a) Comparison of SSE while varying the width $\sigma_B$ of the magnetic field distribution. Negative values $\Delta < 0 $ indicate a single broad distribution is better supported by the evidence.
    (b) Bimodal distribution of magnetic fields chosen for the two-component fit. The widths and positions of the distribution are kept fixed, only the relative proportions of the $B=0$ and  $|B| >0$ are varied.
    }
	\label{fig:residuals}
\end{figure}

For calculation of the $L(\nu)_\mathrm{WFM}$, we still need to include a finite width of the magnetic field distribution to account for variations in nanodiamond orientation, sample-nanodiamond distances, and grain orientations. We calculated $L(\nu)_\mathrm{WFM}$ for several distribution widths $\sigma_{B,WFM}$, fitted the two-component functions to the low-temperature spectrum, and compared the sum of square errors (SSE) of the two-component fit with the SSE of a single distribution fit. We plot the difference $\Delta SSE  = SSE_\mathrm{two-component} - SSE_\mathrm{single~component}$ as a function of the chosen distribution width $\sigma_B$ in Fig. \ref{fig:residuals} (a). Negative values $\Delta \mathrm{SSE} <0 $ indicate a single broad distribution is better supported by the evidence, while $\Delta \mathrm{SSE} >0  $ favors the two-component fit. For calculating the temperature dependence of spectral component ratios, we chose the smallest $\sigma$ for which $\Delta \mathrm{SSE} >0 $, namely $\sigma_B = 0.27 \mathunit{mT}$. The combined, bimodal distribution of magnetic field used in the two-component fitting is shown in Fig \ref{fig:residuals} (b), containing one narrow peak at $B=0$ and a broader peak centered at $|B| = 1.2 \mathunit{mT}$. When fitting the temperature dependence of spectral components, both widths and positions of the distribution are kept fixed and only relative proportions of the $B=0$ and  $|B| >0$ are fitted to the experimental ODMR spectra. \par

We acknowledge that this fitting procedure has limitations arising from the distribution of grain orientations in the compressed powder sample. Since the magnetization direction changes from grain to grain, the average magnetic field magnitude experienced by the NV centers also changes. In this discussion, we simulated the low-temperature spectrum at the location with the largest mean magnetic field and used it for two-component fitting at all measured positions.  However, we have no independent confirmation that our data contain such a measurement; thus, values from such fitting likely overestimate the weakly ferromagnetic spectral ratio.
Accurate determination of the ratio between the antiferromagnetic and weakly ferromagnetic spectral weight would only be possible when complemented with an independent measurement of the magnetic field on the sample surface. 

\end{document}